\documentclass[preprint,12pt]{elsarticle}

\usepackage{graphicx}
\usepackage{amssymb}
 \usepackage{amsthm}
\theoremstyle{definition}
\newtheorem{example}{Example}[section]
\usepackage{pifont}
\usepackage{graphics}
\usepackage{amsmath}
\usepackage{amssymb}
\usepackage{url}
\usepackage{color}
\usepackage{multirow}
\usepackage[table,xcdraw]{xcolor}
\usepackage{colortbl}
\usepackage{epstopdf}

\usepackage[linesnumbered,ruled,vlined]{algorithm2e}

\DeclareMathOperator*{\argmax}{\arg\!\max}
\newcommand{\INV}{\mathbb{INV}}
\newcommand{\SIM}{\mathbb{SIM}}
\newcommand{\ENDISCO}{{\tt EnDisCo}}
\newcommand{\MEDOC}{{\tt MeDOC$++$}}
\newcommand{\LOU}{{\tt Louvain}}
\newcommand{\CNM}{{\tt CNM}}
\newcommand{\INF}{{\tt InfoMap}}
\newcommand{\WT}{{\tt WalkTrap}}
\newcommand{\LP}{{\tt LabelPr}}
\newcommand{\CC}{{\tt ConsCl}}
\newcommand{\FG}{{\tt FstGrdy}}
\newcommand{\OSLOM}{{\tt OSLOM}}
\newcommand{\EAGLE}{{\tt EAGLE}}
\newcommand{\COPRA}{{\tt COPRA}} 
\newcommand{\SLPA}{{\tt SLPA}}
\newcommand{\MOSES}{{\tt MOSES}}
\newcommand{\BIGCLAM}{{\tt  BIGCLAM}}
\newcommand{\MAKGLE}{{\tt M-E}}

\journal{Journal Name}

\begin{document}

\begin{frontmatter}


\title{Ensemble-Based Discovery of Disjoint, Overlapping and Fuzzy Community Structures in Networks}



\author{Tanmoy Chakraborty, Noseong Park}

\address{Indraprastha Institute of Information Technology Delhi (IIIT-D), India\\University of North Carolina, Charlotte, USA}

\begin{abstract}
Though much work has been done on ensemble clustering in data mining, the application of ensemble methods to community detection in networks is in its infancy. In this paper, we propose two ensemble methods: \ENDISCO\ and \MEDOC. \ENDISCO\ performs disjoint community detection. In contrast, \MEDOC\ performs disjoint, overlapping, and fuzzy community detection and represents the first ever ensemble method for fuzzy and overlapping community detection. We run extensive experiments with both algorithms against both synthetic and several real-world datasets for which community structures are known. We show that \ENDISCO\ and \MEDOC\ both beat the best-known existing standalone community detection algorithms (though we emphasize that they leverage them). In the case of disjoint community detection, we show that both \ENDISCO\ and \MEDOC\ beat an existing ensemble community detection algorithm both in terms of multiple accuracy measures and run-time. We further show that our ensemble algorithms can help explore core-periphery structure of network communities, identify stable communities in dynamic networks and help solve the ``degeneracy of solutions'' problem, generating robust results.
\end{abstract}


\end{frontmatter}


\section{Introduction}
Though most human beings recognize a community of people when they see one, coming up with a formal mathematical definition of a community has proved challenging, leading to a plethora of diverse technical definitions which capture the intuitive human understanding of a community with different degrees of accuracy \cite{Fortunato201075}. Because of this, different definitions of a community use different objective functions (such as modularity \cite{Newman:2006:}, significance \cite{TraagKD13}, permanence \cite{Chakraborty:2014,tanmoyjstat,Chakraborty:tkdd}), whose optimization leads to the detection of the underlying community structure. 

We ask ourselves the question: {\em can we come up with a computational model of known communities in a network that accurately approximates real-world communities in the network by leveraging the best known existing network community detection algorithms and their associated objective functions?} We are not the first to examine this question - ensemble methods have already been pioneered in network community detection by 
 \cite{DahlinS13,lanc12consensus}, building on past work on clustering (in non-network settings) in data mining \cite{Xu:2005}. 

Apart from these factors, many other factors suggest that an ensemble-based approach may lead to significantly improved accuracy:

\begin{itemize}
 \item {\bf Dependence on vertex ordering:} Existing community finding algorithms are highly dependent on vertex-ordering \cite{good2010performance,chakraborty}. If we let an algorithm start from different seed vertices in different iterations for a particular network, it might produce completely different community structures.  Each structure can be viewed as a different view of what communities might exist.
 
 \item {\bf Degeneracy of solution:} Existing community finding algorithms suffer from the ``degeneracy of solutions'' problem because they admit an exponential number of distinct high-scoring solutions and typically lack a clear global maximum. Each such solution represents a possible view of the underlying community structure and there is no reason to prefer on over another.
 
 \item {\bf Lack of ground-truth communities:} Most real-world networks do not have (partial) ground-truth community structure to validate predicted communities. Therefore, it is extremely challenging to do any cross-validation to tune the input parameters of an algorithm. 
\end{itemize}

In this paper, we extend our previous study \cite{7752216} where we suggested how to combine multiple solutions to generate ensemble community detection algorithms. However, it was unclear how one can select base solutions. Moreover, there was no single algorithm which can be able to  detect three types of communities  -- disjoint, overlapping and fuzzy. In this paper, we present a comprehensive experiment to show the Superiority of two of our proposed algorithms -- \ENDISCO, and \MEDOC, along with our previous findings \cite{7752216}. In particular, the summary of the contributions presented in this paper is as follows:

\begin{enumerate}
 \item We propose an ensemble-based algorithm called \ENDISCO\ that identifies disjoint communities. \ENDISCO\ is built on the idea that the larger the number of algorithms that place two vertices in the same community, the more likely it is that they really do belong to the same community. We represent each vertex in a feature space and capture the pair-wise distances between vertices. This in turn generates a latent network, capturing the hidden similarities among vertices. A re-clustering algorithm is then  used to cluster the vertices in the latent network to produce the final community structure (Section \ref{sec:endisco}).

 \item We propose \MEDOC, a meta-clustering based algorithm that to the best of our knowledge is the first ensemble-based generalized community detection algorithm to detect disjoint, overlapping and fuzzy communities in a network. The idea behind this algorithm is to generate meta communities  from the ``base communities'' (i.e. communities generated by existing community detection algorithms) by grouping redundant solutions together. We propose a vertex-to-community association function that provides an accurate estimate of the community membership of different vertices in a network. This association function further allows us to detect overlapping and fuzzy  community structures from the network (Section \ref{sec:medoc}).

 \item We conduct an experimental analysis using both synthetic and real-world networks whose ground-truth community structure is known. Experimental results are shown separately for disjoint (Section \ref{sec:disjoint_result}), overlapping (Section \ref{sec:overlapping_result}) and fuzzy (Section \ref{sec:fuzzy_result}) community detection. We show that our ensemble-based algorithms (particularly \MEDOC) outperform all the state-of-the-art baseline algorithms with significant margin including the best known and best performing ``consensus clustering'' algorithm for disjoint community detection \cite{lanc12consensus}. We also present a detailed explanation of how to choose the parameters of the ensemble algorithms.

 \item We provide four strategies to choose a subset of ``base'' community detection algorithms in order to obtain highly accurate communities. These strategies are based on two fundamental quantities -- {\em quality} and {\em diversity}. We show that a trade-off of these two quantities is required to select the best subset of base solutions (Section \ref{base_selection}).
 
 \item \MEDOC~further allows us to explore the core-periphery structure of communities in a network. We observe that vertices with more association within a community form the core unit of the community, whereas peripheral vertices are loosely associated with the community. Furthermore, we show that one can use \MEDOC~to detect communities in a dynamic time-varying network that are stable, i.e., remain almost invariant over time (Section \ref{sec:implication}).
 
 \item Finally we show that both \ENDISCO\ and \MEDOC\ significantly reduce the problem of ``degeneracy of solutions'' in community detection (Section \ref{sec:degeneracy}) and are much faster than consensus clustering \cite{lanc12consensus}, a recently proposed ensemble-based disjoint community finding algorithm (Section \ref{sec:speed}).  
\end{enumerate}

In this paper, the major additions of new contributions with our existing work \cite{7752216} are as follows: (i) We present \MEDOC, the first ensemble based algorithm that can detect disjoint, overlapping and fuzzy community structures (Section \ref{sec:medoc}). (ii) We present results of detecting fuzzy community structures from synthetic and real-world networks (Section \ref{sec:fuzzy_result}). To the best of our knowledge, this is the first time where the fuzzy community detection algorithm is verified against real-world ground-truth. (iii) We present four strategies to select a subset of base solutions which produce near-optimal results (Section \ref{base_selection}). (iv) We present two new implications of \MEDOC~ algorithm -- (a) it can explore the core-periphery structure of communities in a network,  (b) it can detect stable communities in dynamic networks (Section 9). (v)	We present detailed analysis to show how our proposed algorithms reduce the effect of ``degeneracy of solutions'' compared to other baseline algorithms (Section \ref{sec:degeneracy}).

Throughout the paper, we use the term ``community structure'' to indicate the result returned by a community detection algorithm. A community structure therefore consists of a set of ``communities'' and each community consists of a set of vertices.


\section{Related Work}
In this section, we present the literature related to the community detection in three subparts. First, we describe past efforts on disjoint community detection. Second, we discuss fuzzy and overlapping community detection. Finally, we present related work on ensemble-based community detection. Due to the abundance of literature on community detection, we restrict our discussion to best known and/or most recent works. Extensive survey articles on community detection are available in \cite{Fortunato201075}, \cite{chakraborty2016metrics} and \cite{Xie:2013}.

\subsection{Disjoint Community Detection}
Past efforts devoted to community detection mostly assume that nodes are densely connected within a community and sparsely connected across communities. Such efforts include modularity  optimization~\cite{blondel2008,Clauset2004,Guimera,newman03fast,Newman:2006:}, spectral
graph-partitioning~\cite{Newman_13,Thomas}, clique percolation~\cite{Vicsek,PalEtAl05}, local expansion~\cite{Baumes:2005,Lancichinetti},
 random-walk based
approaches~\cite{DeMeo:2013,JGAA-124}, information theoretic approaches~\cite{rosvall2007,Rosvall29012008}, diffusion-based
approaches~\cite{Raghavan-2007}, significance-based approaches~\cite{oslom} and label propagation~\cite{Raghavan-2007,abs-1105-3264,Xie}.
Most of these efforts detect communities in static networks. On the other hand, a large number of algorithms were proposed
to detect communities in dynamically evolving networks (i.e., Internet, Online Social Networks), such as
LabelRankT~\cite{Chen_13}, Estrangement~\cite{Kawadia_NSR2012} and intrinsically dynamic community detection
algorithm~\cite{Bivas_11}. 
Several pre-processing
techniques~\cite{sriram,seed-set-tr} have been developed to improve the quality of the solutions generated by these algorithm. These methods generate preliminary community structure on a set of selected vertices and then modify the structure over successive steps to cover all the vertices. 
Recently, \cite{chakraborty} showed the effect of vertex ordering on the performance of the community detection algorithms.

\subsection{Fuzzy and Overlapping Community Detection}
Another set of community detection algorithms allow a vertex
to be a part of multiple communities.
``CFinder'' \cite{PalEtAl05} was the first method of this kind which was based on clique-percolation technique. However, since more real-world networks are sparse in nature, CFinder generally produces low quality output \cite{Fortunato:2009}. The
idea of partitioning links instead of vertices to discover community structure has also been explored \cite{nature2010}. Some algorithms are based on  
 local expansion and optimization  such as LFM \cite{Lancichinetti}, OSLOM \cite{oslom}, EAGLE \cite{Shen}, MOSES \cite{moses} and GCE \cite{lee}.  \cite{Gregory2011,Zhao} proposed fuzzy community detection technique. 
BIGCLAM \cite{Leskovec} uses Nonnegative Matrix Factorization (NMF) framework for fuzzy/overlapping community detection.   Zhang et al. used NMF to
detect overlapping communities given the feature vector of vertices and known number of communities \cite{11}. 

There is another type of algorithms which exploit  local expansion and optimization to detect overlapping communities. For instance,   ``RankRemoval'' algorithm \cite{BaumesGKMP05} uses a local density function. LFM \cite{Andrea} and MONC \cite{abs-1012-1269} maximize a fitness function over successive iterations. OSLOM \cite{oslom} measures the
statistical significance of a cluster w.r.t a global null model during community expansion. \cite{chen2010detecting} proposed a combination of ``belongingness'' and ``modified modularity'' 
for local expansion. EAGLE \cite{Shen} and GCE \cite{lee} use an agglomerative framework to detection
overlapping communities. COCD \cite{du} first identifies cores after which the remaining vertices are attached to the cores with which they have the largest
connections. 

\cite{14} modeled overlapping community detection as a nonlinear constrained
optimization problem and solved by simulated annealing methods. \cite{Newman05062007,citeulike:4205011,Nowicki,Zareii} used mixer model to solve this problem. \cite{ding} used affinity propagation clustering methods for overlapping community detection. \cite{Whang:2013} proposed a seed set expansion approach for community detection.  

The label propagation algorithm has been extended to overlapping community detection.  COPRA
\cite{Gregory1} updates ``belonging coefficient'' of a vertex by averaging the coefficient from all its neighbors at each time step. SLPA \cite{Xie,abs-1105-3264} propagates labels across vertices based on the pairwise interaction rules. \cite{Chen:2010} proposed a game-theoretic framework in which a community is associated with a Nash local equilibrium.

Beside these, CONGA \cite{Gregory:2007} uses GN algorithm \cite{Girvan2002} to split a vertex  into multiple copies.
\cite{ZhangWWZ09} proposed an iterative process that reinforces the network topology and proximity that is interpreted as the
probability of a pair of vertices belonging to the same community.  \cite{pone.0012528} proposed an approach focusing on
centrality-based influence functions.  
\cite{Sun_11}
proposed fuzzy clustering based disjoint community detection technique.

\subsection{Community Detection using Ensemble Approach} There has been a plethora of research in traditional data mining (not involving networks) to cluster data points using an ensemble approach (see \cite{Xu:2005} for a detailed review). These approaches
can be classified into two categories \cite{Xu:2005}: object co-occurrence based approaches and median partitioning based approaches. However, when it comes to the case of clustering vertices in networks, there are very few such attempts. \cite{DahlinS13} proposed an instance-based ensemble clustering method for network data by fusing different community structures.   \cite{Raghavan-2007} addressed the advantages of combining multiple community structures. CGGC  \cite{OvelgonneG12} presents a modularity maximization based ensemble technique. YASCA is another ensemble approach that can detect ego-centered communities \cite{Kanawati2014,Kanawati14}, and identified the importance of the quality and diversity of base outputs \cite{Kanawati2015}.

Recently, \cite{lanc12consensus} proposed ``consensus clustering'' which leverages a {\em consensus matrix} for disjoint graph clustering. It detects consensus clustering by reweighting the edges based on how many times the pair of vertices are allocated to the same community by different identification methods. This has been proved to be a stable method, outperforming previous approaches for disjoint community detection. However, we differ from them w.r.t. four different points as follows:
\begin{enumerate}
 \item Past work measures the number of times two vertices are assigned to the same community, thus ignoring the global similarity of vertices; whereas we aim at capturing the global aspect by representing the network into a feature space or grouping the redundant base communities into a meta community.
 
\item They either take multiple algorithms or run a particular algorithm multiple times for generating inputs to an ensemble algorithm, whereas we consider both of them.

\item They run algorithms multiple times to generate consensus matrix in each iteration, and keep on repeating the same steps until the matrix converges to a block diagonal matrix which leads to a huge computational cost; whereas we run base algorithms multiple time {\em only} in the first step and leverage the base solutions in later steps which decreases the overall runtime. This leads to significant performance gains.

\item We are the first to show how aggregating multiple disjoint base communities can lead to discover  disjoint, overlapping and fuzzy community structures simultaneously. 
 
\end{enumerate}

However, we consider consensus clustering as one of the state-of-the-art techniques for disjoint community detection and compare it with our algorithms. For overlapping and fuzzy community detection, we present the first ever ``ensemble based'' algorithm in the literature.


\begin{table}[!t]
 \centering
 \caption{Few important notations used in this paper.}\label{notation}
 \scalebox{0.7}{
 \begin{tabular}{c|p{1.1\columnwidth}}
 \hline
 {\bf Notation} & \multicolumn{1}{c}{{\bf Description}} \\\hline
 $G(V,E)$ & An undirected network with sets of vertices $V$ and edges $E$\\ 
 $\mathcal{AL}$ & $\{Al_{m=1}^M\}$, set of $M$ base disjoint community detection algorithms \\ 
 $K$ & Number of iterations (number of vertex orderings) \\ 
$\mathbb{C}_m^k$ & $\{C_m^{1k},...,C_m^{ak}\}$, base community structure discovered by a base disjoint algorithm  $Al_m$ on $k^{th}$ vertex ordering \\ 
 $\Gamma_m$ & $\{\mathbb{C}_m^k\}_{k=1}^K$, set of  base disjoint community structures discovered by base algorithm $Al_m$ on $K$ different vertex orderings\\ 
$\Gamma$ & $\Gamma_{m=1}^M$, set of all $MK$ base disjoint community structures\\
$F(v)$ & Feature vector of a vertex $v$ \\ 
$D_v$ & Maximum distance of $v$ to any community \\ 
 $Clu$& $\bar{a}MK$, approximate total number of communities in $\Gamma$ ($\bar{a}$ is the average size of a base community structure)\\ 
$\mathbb{INV}(v,c)$ & Involvement function of $v$ in community $C$\\ 
$P(C_i|v)$ & Posterior probability of $v$ being part of a community $C_i$\\ 
$P(v)$ & probability distribution of $v$ being part of different communities\\ 
$\mathbb{SIM}(u,v)$ & Similarity function between two vertices \\ 
$\mathbb{M}$ & Ensemble matrix, where $\mathbb{M}(u,v)$ indicates the similarity of vertices $u$ and $v$\\ 
$W(C_i, C_j)$ & Matching function between two communities\\ 
$\mathcal{F}(v,C)$ & A function measuring the association between vertex $v$ and community $C$\\ 
$\mathbb{C}_{GP}$ & $\{C_{GP}^l\}_{l=1}^L$, meta-communities obtained from P-partite graph $GP$\\ 
$\mathbb{A}$ & Association matrix, where $\mathbb{A}(v,l)$ indicates the association of $v$ in meta-community $l$\\ 
$\hat{\mathbb{A}}$ & Normalized association matrix, where $\hat{\mathbb{A}}(v,l)=\frac{\mathbb{A}(v,l)}{\sum_{l'\in L}\mathbb{A}(v,l')}$\\ 
 $\tau$  & A thresholds needed to detect the overlapping community structure in \MEDOC\\ 
 $\mathbb{OC}$ & Final disjoint community structure \\ 
 $\mathbb{OC}$ &  Final overlapping community structure\\ 
 $\mathbb{FC}$ & Final fuzzy community structure\\\hline
 
 \end{tabular}}
\vspace{-5mm}
\end{table}


\begin{algorithm}\small
\caption{\ENDISCO: {\bf En}semble-based {\bf Dis}joint {\bf Co}mmunity Detection }\label{insimul}
\KwData{Graph $G(V,E)$; \\Base algorithms $\mathcal{AL}=\{Al_m\}_{m=1}^M$; \\$K$: Number of iterations; \\$\INV(.,.)$: Involvement function;\\ $\SIM(.,.)$: Similarity function between two vectors; \\$RAlgo$: Algorithm for re-clustering}
\KwResult{Disjoint community structure $\mathbb{DC}$}

$\Gamma=\phi$   \hfill \tcp{Set of all base community structures} 

\tcp{{\color{blue} Generating base partitions}}
\For{each algorithm $Al_m\in \mathcal{AL}$}{
Run $Al_m$ on $G$ for $K$ different vertex orderings and obtain $K$ community structures, denoted by the set $\Gamma_m$; each community structure $\mathbb{C}_m^k \in \Gamma_m$ is of different size and indicated by $\mathbb{C}_m^k=\{C_m^{1k},...,C_m^{ak}\}$;\label{algo1:ensemble}\\
$\Gamma=\Gamma \cup \Gamma_m$;
}

\For{each $v$ in $V$}{\label{algo1:v}
 $F(v)=\phi$; \hfill \tcp{Feature vector of $v$}
 $D_v=0$; \hfill \tcp{Max distance of $v$ to any community}
 $Clu=0$; \hfill \tcp{Total no of communities}
 \tcp{{\color{blue} Constructing ensemble matrix}}
 \For{each $\Gamma_m \in \Gamma$}{\label{gammas} 
     \For{each $\mathbb{C}_m^k\in \Gamma_m$}{
      \For{each $C\in \mathbb{C}_m^k$}{\label{algo1:c}
       Compute $d_v^C=1-\INV(v,C)$;\label{algo1:dist} \\
       $F(v)=F(v) \cup d_v^C$;\label{algo1:fea}\\
       \If{$ d_v^C \geq D_v $}{
	    $D_v=d_v^C$;\label{algo1:D}
        }
        $Clu=Clu+1$;\label{algo1:cl}
       }      
   }
 }\label{gammae} 
 $P(v)=\phi$;\\
 \For{each $F_i(v) \in F(v)$}{\label{algo1:pre}
     \tcp{Posterior probability of $v$ in $C_i^k$}
     Compute $P(C_i|v) =\frac{D_v-F_i(v) +1}{Clu \cdot D_v + Clu -\sum_{k=1}^{Clu} F_k(v)}$;\label{algo1:prob}\\
     $P(v)=P(v)\cup P(C_i|v)$;\label{algo1:post}
   } 
 }
Build an ensemble matrix $\mathbb{M}_{|V|\times|V|}$, where $\forall u,v\in V;\ \mathbb{M}(u,v)$=$\SIM(P(u),P(v))$;\label{algo1:sim}\\
\tcp{{\color{blue} Re-clustering the vertices from $M$}}
Run $RAlgo$ for {\em re-clustering vertices} from $M$ and discover a disjoint community structure $\mathbb{DC}$;\label{algo1:assign} \\
\Return $\mathbb{DC}$
\end{algorithm}


\section{EnDisCo: Ensemble-based Disjoint Community Detection}\label{sec:endisco}
In this section, we present \ENDISCO~({\bf En}semble-based {\bf Dis}joint {\bf Co}mmunity Detection), an ensemble based algorithm to generate disjoint communities in networks.  \ENDISCO\  generates a {\em strong} start by producing different community structures using an ensemble of base community detection algorithms. Then an involvement function is used to measure the extent to which a vertex is involved with  a specific community detected by a base algorithm. This in turn sets the  posterior probability that a vertex belongs to any one of many different communities. This step transforms a network into a feature space. Following this, an ensemble matrix that measures the pair-wise similarity of vertices in the feature space is constructed, and this serves as a latent adjacency matrix in the next step. Finally, we apply a re-clustering algorithm on the ensemble matrix and discover the final disjoint community structure.

\subsection{Algorithmic Description}
\begin{figure}[!t]
 \centering
 \scalebox{0.35}{
 \includegraphics{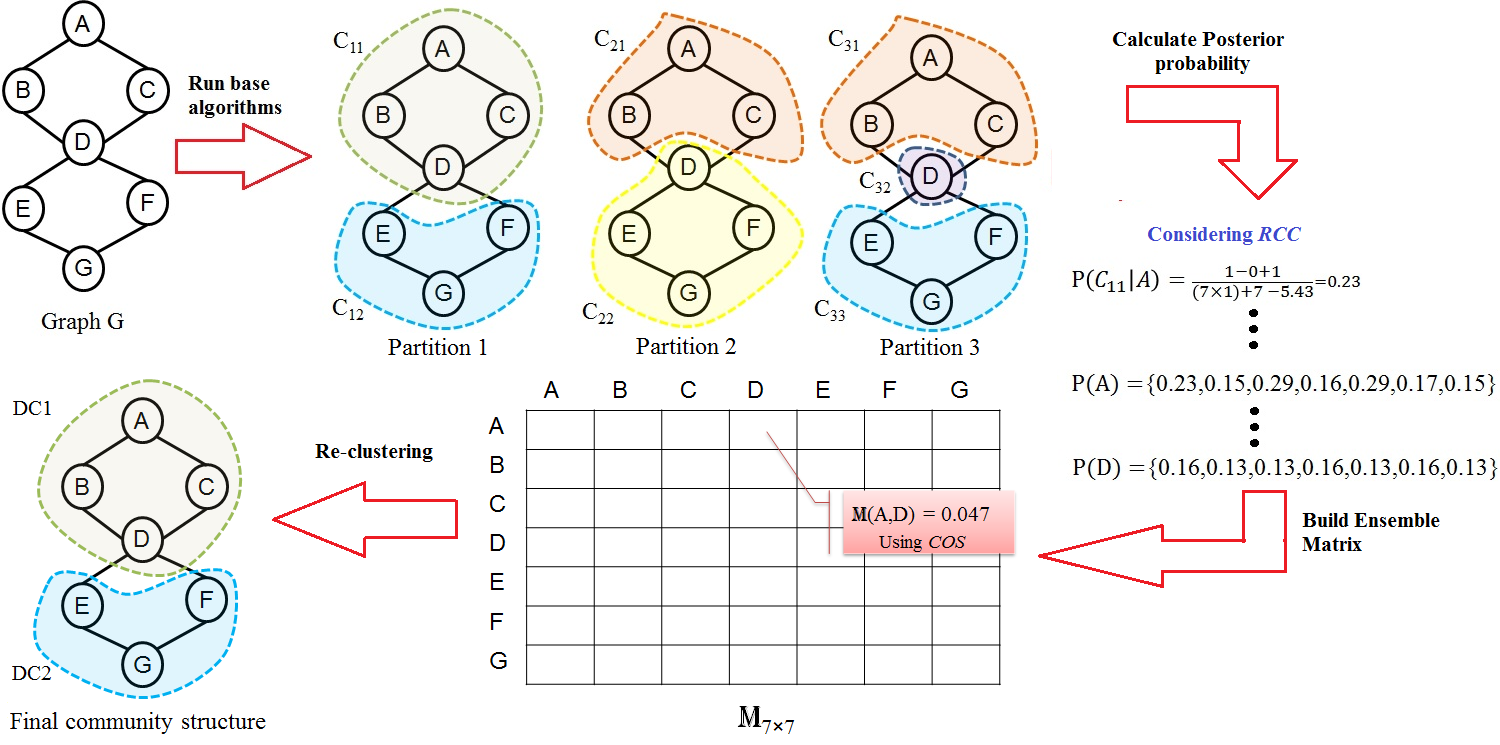}}
 \caption{A toy example depicting the work-flow of \ENDISCO~algorithm. The broken lines indicate the community boundaries. Assume that the base algorithms produce three different community structures in Step 2, and we use Restricted Closeness Centrality ($RCC$) and Cosine Similarity ($COS$) as involvement and similarity functions respectively.  The definitions of $COS$ and $RCC$ are described in Section \ref{algo1:Paramater}.}\label{endisco_demo}
\end{figure}

 \ENDISCO~ follows three fundamental steps (a pseudo-code is shown in Algorithm \ref{insimul}, a toy example of the work-flow is presented in Figure \ref{endisco_demo}, and important notations are shown in Table \ref{notation}):\\

\noindent{\bf (i) Generating base partitions.} 
Given a network $G=(V,E)$ and a set $\mathcal{AL}=\{Al_m\}_{m=1}^M$ of $M$ different base community detection algorithms, \ENDISCO~runs each 
algorithm $Al_m$ on $K$ different vertex orderings (randomly selected) of $G$. This generates a set of $K$ different community structures denoted $\Gamma_m=\{\mathbb{C}_m^k\}_{k=1}^K$, where each community structure $\mathbb{C}_m^k=\{C_m^{1k},\cdots,C_m^{ak}\}$ constitutes a specific partitioning of vertices in $G$, and each $\mathbb{C}_m^k$ might be of different size (Step \ref{algo1:ensemble}). \\

\noindent{\bf (ii) Constructing ensemble matrix.} Given a $\Gamma_m$, we then compute the extent of $v$'s involvement in each community $C$ in $\mathbb{C}_m^k$ via an ``involvement'' function $\INV(v,C)$ (Step \ref{algo1:dist}). Possible definitions of $\INV$ are given in Section \ref{algo1:Paramater}. For  each vertex $v$, we construct a feature vector $F(v)$ whose elements indicate the distance  of $v$ (measured by $1-\INV$) from each community obtained from different runs of the base algorithms (Step \ref{algo1:fea}). The size of $F(v)$ is the number of communities $Clu$ in $\Gamma$ (approx. $\bar{a}MK$, where $\bar{a}$ is the average size of a base community structure). Let $D_v$ be the largest distance of $v$ from any community in the sets in $\Gamma$ (i.e., $D_v=\max_i F_i(v)$ in Step \ref{algo1:D}). We define the conditional probability of $v$ belonging to community $C_i$  (Step \ref{algo1:prob}) as:
\begin{equation}\small
 P(C_i|v) =\frac{D_v-F_i(v) +1}{Clu \cdot D_v + Clu -\sum_{k=1}^{Clu} F_k(v)}
\end{equation}
The numerator ensures that the greater the distance $F_i(v)$ of $v$ from community $C_i$, the less likely $v$ is to be in community $C_i$. 
The normalization factor in the denominator ensures that $\sum_{k=1}^{Clu} P(C_i|v)=1$. We further observe that $\forall v\in V$ and $\forall i$, $P(C_i|v) >0$. 
Add-one smoothing in the numerator allows a non-zero probability to be assigned to all $C_i$s, especially for $C_{\hat{k}}$ such that $\hat{k}=\argmax\limits_{k} F_k(v)$. In this case $P(C_i|v)$ is assigned its maximum value $P(C_k|v)=\frac{1}{Clu \cdot D_v + Clu -\sum_{k=1}^{Clu} F_k(v)}$. 
The larger the deviation of $F_k(v)$ from $D_v$, the more the increase of $P(C_i|v)$, the corresponding community $C_i$ becomes more likely for $v$.

The set of posterior probabilities of $v$ is: $P(v)=\{ P(C_k|v) \}_{k=1}^ {Clu}$ (Step \ref{algo1:post}), which in turn transforms a vertex into a point in a multi-dimensional feature space. Finally, we construct an ensemble matrix $M$ whose entry $M(u,v)$ is the similarity (obtained from a function $\SIM$ whose possible definitions are given in Section \ref{algo1:Paramater}) between the feature vectors of $u$ and $v$ (Step \ref{algo1:sim}). The ensemble matrix ensures that the more communities a pair of vertices share the more likely they are connected in the network \cite{Leskovec}. \\

\noindent{\bf (iii) Discovering final community structure.}  In Step \ref{algo1:assign} we use a community detection algorithm $RAlgo$ to re-cluster the vertices from $M$ and discover the final disjoint community structure (Step \ref{algo1:assign}).

\subsection{Parameter Selection}\label{algo1:Paramater}
 We now describe different parameters of \ENDISCO:
 
 \begin{itemize}
  \item  {\bf Involvement Function ($\INV$):} We use two functions to measure the involvement of a vertex $v$ in a community $C$: (i) {\em Restricted Closeness Centrality} ($RCC$): This is the inverse of the average shortest-path distance from the vertex $v$ to the vertices in community $C$, i.e., $RCC(v,C)=\frac{|C|}{\sum_{u\in C} dist(v,u)}$; (ii) {\em Inverse Distance from Centroid} ({\em IDC}): we first identify the vertex with highest closeness centrality (w.r.t. the induced subgraph of $C$) in community $C$, mark it as the centroid of $C$ (denoted by $u_c$), and then measure  the involvement of $v$ as the inverse of the shortest-path distance between $v$ and $u_c$, i.e.,  $IDC(v,C)=\frac{1}{dist(v,u_c)}$.\\

\begin{example}
 In Figure \ref{endisco_demo}, the $RCC$ of vertex $D$ for community $C_{12}$ is measured as follows. The shortest-path distances of $D$ from $E$, $F$ and $G$ are $1$, $1$ and $2$ respectively. Then $RCC(D,C_{12})=\frac{3}{1+1+2}=\frac{3}{4}$. On the other hand, the centroid of community $C_{12}$ is vertex $G$ and the distance between $D$ and $G$ is $2$. Therefore, $IDC(D,C_{12})=\frac{1}{2}$.
\end{example}

\item  {\bf Similarity Function ($\SIM$):} We consider cosine similarity ($COS$) and Chebyshev similarity ($CHE$) (Chebyshev similarity is defined as ($1-CHE_d$) where $CHE_d$ is the Chebyshev distance) to measure the similarity between two vectors.\\

\item  {\bf Algorithm for Re-clustering ($RAlgo$):} we consider each base community detection algorithm as a candidate to re-cluster vertices from the ensemble matrix. The idea is to show that existing community detection algorithms can perform even better when they consider the ensemble matrix of network $G$ as opposed to the adjacency matrix of $G$.  However, one can use any community detection algorithm in this step to detect the community structure. We will show the effect of different algorithms used in this step in Section \ref{sec:impact}.\\

\item {\bf Number of Iterations ($K$):} Instead of fixing a hard value, we set $K$ to be dependent on the number of vertices $|V|$ in the network. We vary $K$ from $0.01$ to $0.50$ (with step $0.05$) of $|V|$ and confirm that for most of the networks, the accuracy of the algorithm converges at $K=0.2|V|$ (Figures \ref{parameter_dis}(c) and \ref{parameter_dis}(f)), and therefore we set $K=0.2|V|$
in our experiments. 

 \end{itemize}

\subsection{Complexity Analysis}\label{algo1:complexity}

Suppose $N=|V|$ is the number of vertices in the network, $M$ is the  number of base algorithms and $K$ is the number of vertex orderings. Further suppose $\bar{a}$ is the average size of the community structure. Then the loop in Step \ref{algo1:v} of Algorithm 1  would iterate $\bar{a}NMK$ times (where $M,K\ll N$). The construction of the ensemble matrix in Step \ref{algo1:sim} would take $\mathcal{O}(N^2)$. 
Graph partitioning is NP-hard even to find a solution with guaranteed approximation bounds --- however, heuristics such as the famous Kernighan-Lin algorithm take $O(N^2\cdot \mbox{log}(N))$ time.

\begin{algorithm}
\caption{{\MEDOC}: A {\bf Me}ta Clustering based {\bf D}isjoint, {\bf O}verlapping and Fu{\bf z}zy Community Detection}\label{meclud}
\KwData{Graph $G(V,E)$;\\ Base algorithms $\mathcal{AL}=\{Al_m\}_{m=1}^ M$;\\ $K$: Number of iterations; \\$W(.,.)$: Matching between pair-wise communities;\\ $RAlgo$: Algorithm for re-clustering; \\$\mathcal{F}(.,.)$: vertex-to-community association; \\ $\tau$: threshold for overlapping community detection}
\KwResult{Disjoint ($\mathbb{DC}$), overlapping ($\mathbb{OC}$) and fuzzy ($\mathbb{FC}$) community structures}
\tcp{{\color{blue} Constructing multipartite network}}
\For{$Al_m$ in $\mathcal{AL}$}{
Run $Al_m$ on $G$ for $K$ different vertex orderings and obtain $K$  community structures, denoted by the set $\Gamma_m=\{\mathbb{C}_m^k\}_{k=1}^K$; each community structure $\mathbb{C}_m^k \in \Gamma_m$ may be of different size and is denoted by $\mathbb{C}_m^k=\{C_m^{1k},...,C_i^{ak}\}$; \label{algo2:perm}}
Construct a $P$-partite graph $GP$ (where $P=M.K$)  consisting of $M.K$ partitions, each corresponding to each community structure obtained in Step 2: vertices in partition $m^k$ are communities in $\mathbb{C}_m^k$ and edges are drawn between two pair-wise vertices (communities) $C_m^{ik}$ and $C_n^{jk}$ with the edge weight $W(C_m^{ik},C_n^{jk'})$;\label{algo2:cons}\\
\tcp{{\color{blue} Re-clustering the multipartite network}}
Run $RAlgo$ to re-cluster vertices in $GP$ and discover a meta-community structure, $\mathbb{C}_{GP}=\{C_{GP}^{l}\}_{l=1}^L$;\label{algo2:run}\\
\tcp{{\color{blue} Constructing an association matrix}}
Construct an association matrix $\mathbb{A}_{|V|\times L}$, where ${\mathbb A}(v,l)=\mathcal{F}(v,C_{GP}^l)$, indicating the association of vertex $v$ to a meta-community $C_{GP}^l$;\label{algo2:asso}\\
\tcp{{\color{blue} Discovering final community structure}}
Each row in $\mathbb{A}$ indicates the memberships of the corresponding vertex in $L$ meta-communities;\\
To get $\mathbb{DC}$, we assign a vertex $v$ to community $C^* = \argmax\limits_{C} \ \mathbb{A} (v,C) $;\label{algo2:dc}\\
To get $\mathbb{OC}$, we assign a vertex $v$ to a set of communities $C_v^*$ so that $\forall C\in C_v^*: \mathbb{A} (v,C) \geq \tau$;\label{algo2:oc}\\
To get $\mathbb{FC}$, we first normalize each entry in $\mathbb{A}$ by the sum of entries in the corresponding row and obtain a normalized association matrix $\hat{\mathbb{A}}$, i.e., $\hat{\mathbb{A}}(v,l)=\frac{\mathbb{A}(v,l)}{\sum_{l'\in L}\mathbb{A}(v,l')}$; and  assign a vertex $v$ to a community $C$ with the membership probability of $\hat{\mathbb{A}} (v,C)$;\label{fuzzy}\\
\Return $\mathbb{DC}$, $\mathbb{OC}$, $\mathbb{FC}$
\end{algorithm}

\begin{figure}[!t]
 \centering
 \scalebox{0.29}{
 \includegraphics{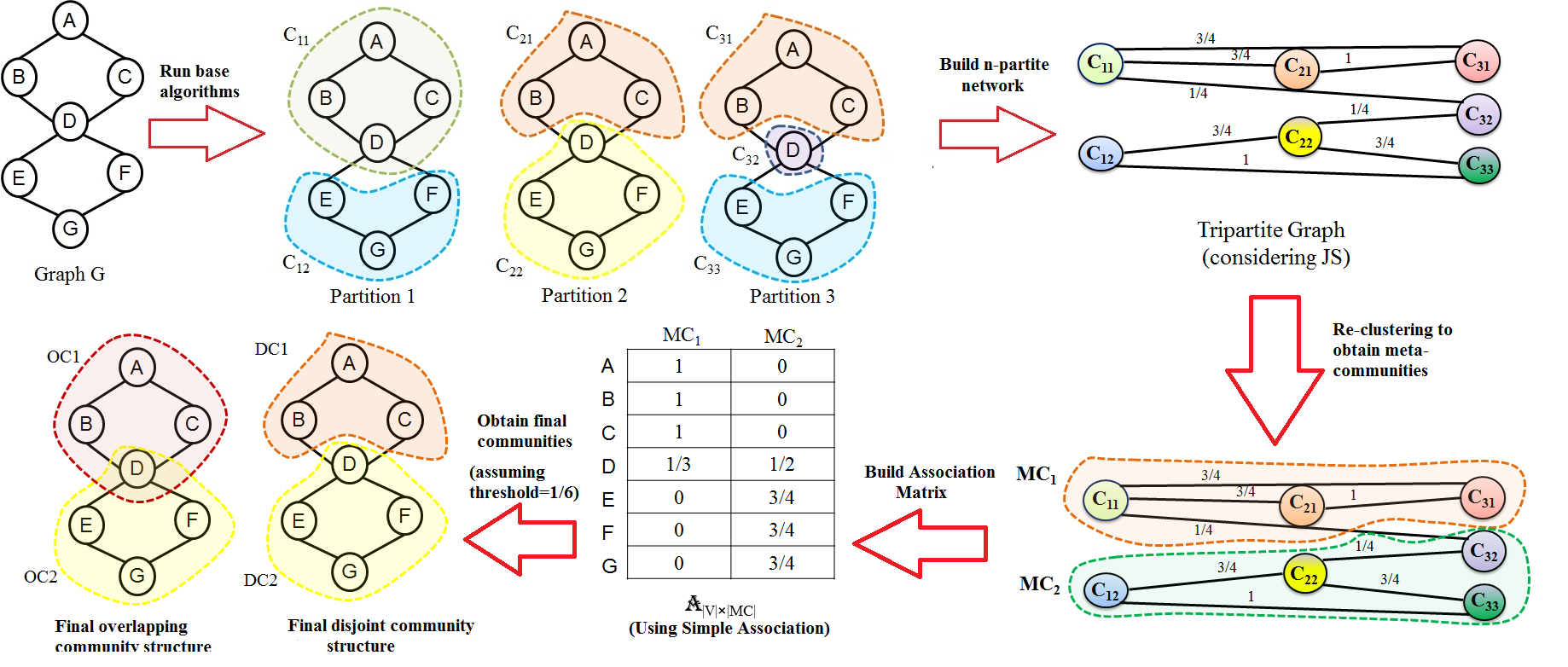}}
 \caption{A toy example depicting the work-flow of \MEDOC~algorithm. The broken lines indicate the community boundaries. Assume that the base algorithms produce three different community structures in Step 2, and we use Jaccard Similarity ($JC$) and simple association ($\mathcal{F}$) as matching and association functions respectively.  The definitions of $JC$ and $\mathcal{F}$ are described in Section \ref{algo2:Parameter}. The threshold $\tau$ is chosen as $\frac{1}{6}$. }\label{Medoc_demo}
\end{figure}

\section{MEDOC$++$: Meta-clustering Approach}\label{sec:medoc}
\MEDOC~  ({\bf Me}ta Clustering based {\bf D}isjoint, {\bf O}verlapping and Fu{\bf z}zy Community Detection) starts by executing all base community detection algorithms, each with different vertex orderings, to generate a set of community structures. It then creates a multipartite
network. After this, another community detection algorithm is used to partition the multipartite network. Finally, a vertex-to-community association function is used to determine the strength of membership of a vertex in a community. Unlike \ENDISCO, \MEDOC~ can yield disjoint, overlapping and fuzzy community structures from the network. 

\subsection{Algorithmic Description}
\MEDOC~ has the following four basic steps (pseudo-code is in Algorithm \ref{meclud}, a toy example of the work-flow of \MEDOC~ presented in Figure \ref{Medoc_demo}, and important notations are shown in Table \ref{notation}):\\

\noindent{\bf (i) Constructing multipartite network.} \MEDOC~takes 
a set $\mathcal{AL}=\{Al_m\}_{m=1}^M$  of 
$M$ base community detection algorithms as input and runs each $Al_m$ on $K$ different vertex orderings of $G$. For each ordering $k$, $Al_m$ produces a community structure $\mathbb{C}_m^k=\{C_m^{1k},...,C_i^{ak}\}$ of varying size (Step \ref{algo2:perm}). After running on $K$ vertex orderings, each algorithm $Al_m$ produces $K$ different community structures $\Gamma_m=\{\mathbb{C}_m^k\}_{k=1}^K$.  Therefore at the end of Step \ref{algo2:perm}, we obtain $K$ community structures each from $M$ algorithms (essentially, we have $P=M.K$ community structures). We now construct a $P$-partite network (aka meta-network) $GP$ as follows:
vertices are members of $\bigcup_{m} \mathbb{C}^k_m$, i.e., a community present in a base community structure (obtained from any of the base algorithms in $\mathcal{ AL}$ and any vertex ordering) is a vertex of $GP$.
We draw an edge from a community $C_m^{ik}$ to a community $C_n^{jk'}$ and associate a weight $W(C_m^{ik},C_n^{jk'})$  (Step \ref{algo2:cons}). Possible definitions of $W$ will be given later in Section~\ref{algo2:Parameter}.
Since each $\mathbb{C}_m^k$ is disjoint, the  vertices in each partition are never connected.   \\

\noindent{\bf (ii) Re-clustering the multipartite network.} 
Here we run any standard community detection algorithm $RAlgo$ on the multipartite network $GP$ and obtain a community structure containing (say) $L$ communities $\mathbb{C}_{GP}=\{C_{GP}^l\}_{l=1}^L$. Note that in this step, we cluster the communities obtained earlier in Step 2; therefore each such community $C_{GP}^l$ obtained here is called a ``meta-community'' (or community of communities) (Step \ref{algo2:run}). \\  

\noindent{\bf (iii) Constructing an association matrix.} We determine the association between a vertex $v$ and a meta-community $C_{GP}^l$ by using a function $\mathcal{F}$  and construct an association matrix $\mathbb{A}$ of size $|V|\times L$, where each entry ${\mathbb A}(v,l)=\mathcal{F}(v,C_{GP}^l)$ (Step \ref{algo2:asso}). Possible definitions of $\mathcal{F}$ will be given later in Section~\ref{algo2:Parameter}.\\

\noindent{\bf (iv) Discovering final community structure.} Final vertex-to-community assignment is performed by processing $\mathbb{A}$. The entries in each row of $A$ denote membership probabilities of the corresponding vertex in $L$ communities. For disjoint community assignment, we label each vertex $v$ by the community $l$ in which $v$ possesses the most probable membership in $\mathbb{A}$, i.e.,  $l^* = \argmax\limits_{l} \ \mathbb{A} (v,l)$. Tie-breaking is handled by assigning the vertex to the community to which most of its direct neighbors belong. Note that not every meta-community can be guaranteed to contain at least one vertex. Thus, we cannot guarantee that there will be $L$ communities in the final community structure. For discovering overlapping community structure, we assign a vertex $v$ to those communities for which the membership probability exceeds a threshold $\tau$. Possible ways to specify this threshold will be specified later in Section~\ref{algo2:Parameter}. 

For fuzzy community detection, we first normalize each entry of $\mathbb{A}(v,l)$ by the sum of entries in the corresponding row so that the membership probability of each vertex in different communities sums up to $1$. This in turn returns a new association matrix $\hat{\mathbb{A}(v,l)}$. Accordingly we assign each vertex $v$ to a community $C$ with the membership probability of $\hat{\mathbb{A}} (v,C)$ (see Step \ref{fuzzy}).

\subsection{Parameter Selection}\label{algo2:Parameter}
Here we describe the parameters used in \MEDOC~algorithm:

\begin{itemize}
\item {\bf Matching Function ($W$)}: Given two communities $C_i$ and $C_j$, we measure their matching/similarity via Jaccard Coefficient ({\em JC})=$\frac{|C_i \cap C_j|}{|C_i \cup C_j|}$ and average precision ({\em AP}) =$\frac{1}{2}(\frac{|C_i \cap C_j|}{|C_i|} + \frac{|C_i \cap C_j|}{|C_j|})$.

\begin{example}
In Figure \ref{Medoc_demo},  the Jaccard Coefficient between $C_{11}$ and $C_{21}$ is $JC(C_{11},C_{21})=\frac{3}{4}$. The average precision between them is $AP(C_{11},C_{21})=\frac{1}{2} (\frac{3}{4} + \frac{3}{3})=\frac{7}{8}$. 
\end{example}

 \item {\bf Association Function ($\mathcal{F}$)}: Given a meta-community $C$ consisting of (say,) $\gamma$ communities, the association of $v$ with $C$ can be calculated as $\mathcal{F}(v,C)=\frac{\sum_{l=1}^{\gamma} \delta(v,C^l)}{\gamma}$, where $\delta$ returns $1$ if $v$ is a part of $C^l$, $0$ otherwise.  Alternatively,
 a weighted association measure may assign a score to $v$ w.r.t. $C$ based on the co-occurrence of the other community members with $v$, i.e., {\scriptsize $\mathcal{F}_w(v,C)=  
\frac{\big|\bigcap \limits_{C^l\in C}  C^l \delta(v,C^l) \big|} {\big|\bigcup\limits_{C^l\in C} C^l \delta(v,C^l)\big|}$}.

\begin{example}
In Figure \ref{Medoc_demo}, the simple association between vertex $A$ and meta community $MC_1$ is $\mathcal{F}(A,MC_1)=\frac{3}{3}=1$ because $A$ is present in $C_{11}$, $C_{21}$ and $C_{31}$ communities which are parts of $MC_1$. On the other hand, the weighted association between $A$ and $MC_1$ is    $\mathcal{F}_w(A,MC_1) = \frac{|\{A,B,C,D\}  \cap \{A,B,C\} \cap \{A,B,C\}|}{|\{A,B,C,D\}  \cup \{A,B,C\} \cup \{A,B,C\}|}=\frac{3}{4}$. 
 
\end{example}

\item {\bf Threshold ($\tau$):} We choose the threshold $\tau$ automatically as follows. We first assign each vertex to its most probable community -- this produces a disjoint community structure. 
Each vertex $v_i$ is represented by a feature vector $F(v_i)$ which is the entire $i$'th row of the association matrix $\mathbb{A}$.
We then measure the average similarity of vertices in $C$ as follows: $AS(C)=\frac{\sum_{(u,v)| u,v \in C \wedge  E_{uv}\in E_C} COS(F(u),F(v))}{|E_C|}$, where $E_C$ is the set of edges completely internal to $C$, $E_{uv}$ is an edge $(u,v)$, and $COS$ is cosine similarity. The probability that two vertices are connected in $C$ is then defined as:
\begin{equation}\small
 P(C)=\frac{e^{{[AS(C)]}^2}}{1+e^{{[AS(C)]}^2}}
\end{equation}
For a vertex $v$, if $P(C\cup \{v\}) \geq P(C)$, we further assign $v$ to $C$, in addition to  its current community.  

\begin{example}
 In Figure \ref{Medoc_demo}, let us measure the threshold for community $DC1$ that we obtain in the final step after discovering the disjoint community structure. From the association matrix $\mathbb{A}$ in the figure, $F(A)=\{1,0\}$, $F(B)=\{1,0\}$ and $F(C)=\{1,0\}$. So the average similarity of vertices in $DC1$ in terms of cosine similarity is $AS(DC1)=1$. Then the probability of two vertices in $DC1$ being connected is $P(DC1)=0.88$. If we add $D$ to $DC1$, the new probability $P(DC1 \cup D)=0.80$ and $P(DC1 \cup D)<P(DC1)$. Therefore, $D$ is not assigned to $DC1$. 
\end{example}

It is worth noting that the selection of a threshold depends upon which community we start with. For instance, in Figure \ref{Medoc_demo}, if we start from community $DC2$ and calculate the increase in probability after assigning $B$ to it, it would treat this assignment as a valid assignment. If we continue assigning the other vertices, the entire network would get assigned to a single community. To avoid this situation, we start from that community for which the membership probability $P(C)$ is highest among all and keep assigning other vertices to it. In Figure \ref{Medoc_demo}, we start from $DC1$. Once the members in a given community are finalized, we will not perturb their community membership further. This in turn also reduces runtime.

We compare our threshold selection method with the following method: each vertex is assigned to its top $n\%$ high probable communities (we set $n$ to $5\%$ or $10\%$). Our experiments show that \MEDOC~delivers excellent performance with our threshold selection method (see Figures \ref{parameter_over}(g) and \ref{parameter_over}(i)). 

\end{itemize}

Other input parameters $RAlgo$ and $K$ remain same as discussed in Section \ref{algo1:Paramater}.

\subsection{Complexity Analysis}
 The most expensive step of \MEDOC~ is to construct the multipartite network in Step 3. If $M$ is the number of base algorithms, $K$ is the number of vertex orderings and $\bar{a}$ is the average size of a base community structure,  the worst case scenario occurs when each vertex in one partition is connected to each vertex in other partitions --- if this happens, the total number of edges is $\mathcal{O}(\bar{a}^2M^2K^2)$. 
 However, in practice the network is extremely sparse and  leads to  $\mathcal{O}(\bar{a}MK)$ edges (because in sparse graphs $\mathcal{O}(|V|)\sim \mathcal{O}(|E|)$).
 Further, constructing the association matrix would take $\mathcal{O}(NL)$ iterations (where $L\ll N$).


\begin{table}
\caption{Properties of the real-world networks with disjoint community structure. $N$: number of vertices, $E$: number of edges, $C$: number of
communities, $\rho$: average edge-density per community, $S$: average size of a community.}\label{dataset_disjoint}
\centering
\scalebox{0.85}{
\begin{tabular}{l||r|r|r|r|r}
\hline
Network &  N & E & C & $\rho$ & $S$ \\\hline
University   &  81      &  817      &    3    & 0.54    &   27   \\
Football  & 115 & 613 & 12 & 0.64  & 9.66  \\
Railway  & 301 & 1,224 & 21 & 0.24 & 13.26   \\  
Coauthorship  & 103,677 & 352,183 & 24 &  0.14  & 3762.58   \\\hline

\end{tabular}}
\end{table}

%
%
%
%

\section{Experiments: Results of Disjoint Community Detection}\label{sec:disjoint_result}
In this section, we evaluate \ENDISCO\ and \MEDOC\ for disjoint community detection. We start by explaining the datasets used in this experiment, followed by the baseline algorithms taken to compare against our algorithms, and the evaluation metrics used to compare the detected communities with the ground-truth. Then in the experimental results we will show how we select the parameters and the comparative analysis.

\subsection{Datasets}\label{dis_dataset}
We use both synthetic networks with community structures embedded, and real-world networks of different sizes with known ground-truth community structure.

\subsubsection{Synthetic Networks}
We use the LFR benchmark model \cite{PhysRevE} to generate synthetic networks with ground-truth community structure by varying the number of vertices $n$, mixing parameter $\mu$ (the ratio of inter- and intra-community edges), average degree $\bar k$, maximum degree $k_{max}$, minimum (maximum) community size $c_{min}$ ($c_{max}$), average percentage $O_n$ of overlapping vertices and the average number $O_m$ of communities to which a vertex belongs. The parameter $\mu$ controls the quality of the community structure -- the more the value of $\mu$, the more the inter-community edges and the less the quality of the community structure. We vary this parameter in order to generate different network and community structures. We also vary $O_m$ and $O_n$ to  obtain communities in different extent of overlapping (see Section \ref{sec:overlapping_result}). Unless otherwise stated, we generate networks with the same parameter configuration used in \cite{Chakraborty:2014,Kanawati2014} for disjoint community structure: $n=10000$, $\bar k=50$, $k_{max}=150$, $\mu=0.3$, $O_n=0$, $O_m=1$, $c_{max}=100$, $c_{min}=20$. 
Note that for each parameter configuration,
we generate 50 LFR networks, and the values in all the experiments are reported by averaging the results.

\subsubsection{Real-world Networks}

We also use the following four real-world networks mentioned in Table \ref{dataset_disjoint} for experiments:

\textbf{Football network:} This network constructed from \cite{Clauset2004} contains the network of American football games between Division IA colleges during the regular season of Fall 2000. The vertices in the network represent teams (identified by their college names) and edges represent regular-season games between the two teams they connect. The teams are divided into conferences (indicating communities) containing around 8-12 teams each. 

\textbf{Railway network:} This network proposed by  \cite{Chakraborty:2014} consists of vertices representing railway stations in India, where two stations $s_i$ and $s_j$ are connected by an edge if there exists at least one train-route such that both $s_i$ and $s_j$ are scheduled halts on that route.  Here the communities are states/provinces of India since the number of trains within each state is much higher than the trains in-between
two states.

\textbf{University network:} This network generated by  \cite{14} is a friendship network of a faculty of a UK university, consisting of 81 vertices (individuals) and 817  connections. The school affiliation of each individual is stored as a vertex attribute. Schools act as communities in this network. 

\textbf{Coauthorship network:}\label{coau}
This network suggested by Chakraborty et al. \cite{Chakraborty:2014}  is derived from the citation dataset \cite{ChakrabortySTGM13}.
Here each vertex represents an author. An undirected edge between authors is drawn if the two authors
coauthor at least one paper. The communities are marked by the research fields since authors have a tendency to collaborate with other authors within the same field. It may be possible that an author has worked on multiple fields, which causes the communities to overlap. We assign an author to that research community in which he/she has published the most papers. However, later in Section \ref{sec:fuzzy_result} for fuzzy community detection we will leverage this information to prepare the fuzzy ground-truth community structure.  

\subsection{Baseline Algorithms}
There exist numerous community detection algorithms which differ in the way they define community structure. Here we select the following set of algorithms as our baselines and categorize them according to the principle they use to identify communities as per \cite{1742-5468-2012-08-P08001}: 
(i) {\bf Modularity-based approaches}: FastGreedy (\FG) \cite{newman03fast}, Louvain (\LOU) \cite{blondel2008} and \CNM~ \cite{Clauset2004};
(ii) {\bf Vertex similarity-based approaches}: WalkTrap (\WT) \cite{JGAA-124};
(iii) {\bf Compression-based approaches}: 
InfoMap (\INF) \cite{Rosvall29012008};
(iv) {\bf Diffusion-based approaches}: Label Propagation (\LP) \cite{Raghavan-2007};
(v) {\bf Ensemble approaches}: The most recent ensemble-based disjoint community detection algorithm is Consensus Clustering (\CC) \cite{lanc12consensus}. This algorithm starts by running a base algorithm multiple times and generates a consensus matrix, i.e., a matrix based on the co-occurrence of vertices in communities of the input partitions. The consensus matrix is further used as an input to the base algorithm adopted, leading to a new set of partitions, which generate a new consensus matrix, until a unique partition is finally reached, which cannot be altered by further iterations.     


Note that all these algorithms, except consensus clustering are also used as base algorithms in $\mathcal{AL}$ in our ensemble approaches. 

\subsection{Evaluation Metrics} \label{dis_eval}
Since the ground-truth community structure is available for each network, we use the following two standard metrics to compare the detected community structure with the ground-truth: Normalized Mutual Information (NMI) \cite{danon2005ccs} and Adjusted Rand Index (ARI) \cite{hubert1985}. The larger the value of NMI and ARI, the better the matching between two community structures.

\if{0}
As we know the ground-truth community structure, a stronger test of the correctness of the community detection algorithm, however, is by comparing the obtained community with a given ground-truth structure. We use two
standard validation metrics as follows:

\begin{itemize}
 \item {\bf Normalized Mutual Information} ({\em NMI}):  
 Given a network $G(V,E)$,  $\Omega=\{\omega_1,\omega_2,\cdots,\omega_K\}$ is the set of detected communities, and $C=\{c_1,c_2,\cdots,c_J\}$ is the set of ground-truth communities. $N=|V|=\sum_{k\in K} |\omega_k|=\sum_{j\in J} |c_j|$ is the total number of nodes, $N_{c_j}=|c_j|$ and $N_{\omega_i c_j}=|\omega_i \cap c_j|$.
 
NMI \cite{danon2005ccs} is a widely-used information-theoretic metric to compare two community structures. It is defined as follows:
\begin{equation}\label{nmi}
NMI(\Omega,C)=\frac{I(\sigma,C)}{[H(\sigma)+H(C)]/2}
\end{equation}
where $I$ is mutual information,
\begin{equation}\label{i}
I(\Omega,C)=\sum_k\sum_j \frac{|\omega_k \cap c_j|}{N}\ log\frac{{N|\omega_k \cap c_j|}}{|\omega_k||c_j|}
\end{equation}

$H$ is entropy as defined below,
\begin{equation}\label{h}
H(\Omega)=-\sum_k \frac{|\omega_k|}{N}\ log\frac{\omega_k}{N}
\end{equation}

NMI is always a number between 0 (no matching) and 1 (perfect matching).

\item {\bf Adjusted Rand Index} ({\em ARI}): In the domain of community detection,  ARI \cite{hubert1985} is often preferred. It seems to be less sensitive to the number of communities.  It follows a general formula applicable for any measure $M$. The chance-corrected version of $M$ (subtracting the random probability from the original value), can be written as follows: 
\begin{equation}\label{m}
 M_c=\frac{M-E(M)}{M_{max}-E(M)}
\end{equation}
where $M_c$ is the chance-corrected version, $M_{max}$ is the maximal value $M$ can reach, and $E(M)$ is the value expected for some null model. Under the assumption that the partitions are generated randomly with the constraint of having fixed number of communities and part sizes, the expected value for the number of pairs in a community intersection $\omega_i \cap c_j$:
\begin{equation}
 E\dbinom{N_{\omega_ic_j}}{2}=\dbinom{N_{\omega_i}}{2}\dbinom{N_{c_j}}{2}/\dbinom{N}{2}
\end{equation}
By replacing in Equation \ref{m} and after some simplifications, we get the final ARI,
\begin{equation}\label{ari}\scriptsize
 ARI(\Omega,C)=\frac{\sum_{ij}\dbinom{N_{\omega_ic_j}}{2} - \sum_i\dbinom{N_{\omega_i}}{2} \sum_j \dbinom{N_{c_j}}{2} / \dbinom{N}{2}}{\frac{1}{2} \bigg(\sum_i\dbinom{N_{\omega_i}}{2} + \sum_j \dbinom{N_{c_j}}{2} \bigg) - \sum_i\dbinom{N_{\omega_i}}{2} \sum_j \dbinom{N_{c_j}}{2} / \dbinom{N}{2}}
\end{equation} 

\end{itemize}

\fi
  
\begin{figure}[!t]
\centering
 \scalebox{0.3}{
 \centering
  \includegraphics{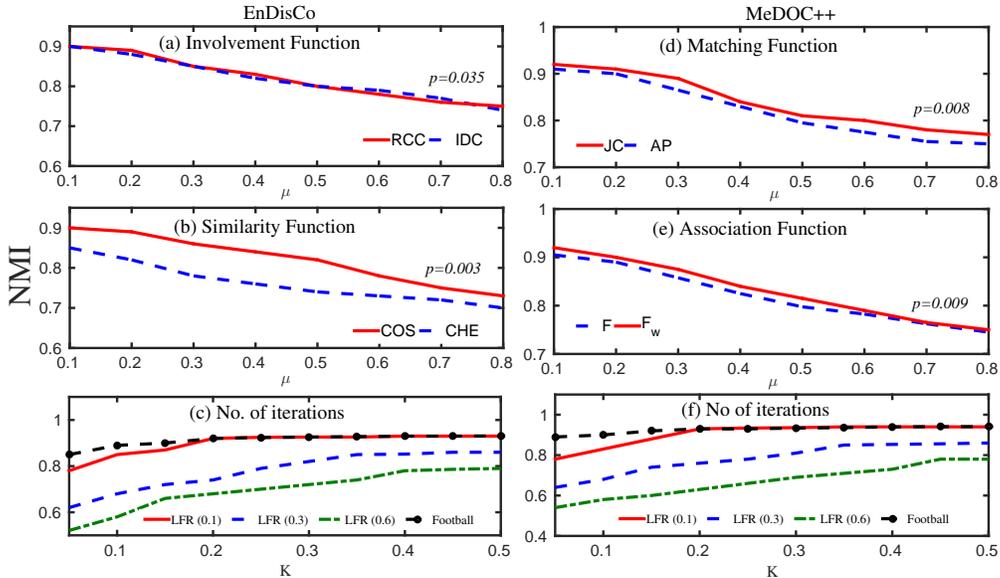}}    
  \caption{Dependencies of the performance of \ENDISCO~ (left panel) and \MEDOC~ (right panel) on different parameters. The quality of the ground-truth community is varied by changing $\mu$ from $0.1$ to $0.8$ (keeping the other LFR parameters default) and the performance is measured using NMI. In (c) and (f), we vary $K$ and report the accuracy for three different LFR and Football networks (the results are similar for the other real networks and are not shown). The value corresponding to one parameter is reported by averaging the values for all possible combinations of the other parameters. The results are statistically significant (for multiple curves in (c) and (f), we report the range of $p$-values).}\label{parameter_dis}
   
\end{figure}

\subsection{Experimental Results}
We first run experiments to identify the best parameters for \ENDISCO~and \MEDOC~ for disjoint community detection and then compare them with competing algorithms.

\subsubsection{Dependency on the Parameters}
We consider the LFR networks and vary  $\mu$. Figure \ref{parameter_dis}(a) shows that the accuracy of \ENDISCO~is similar for both the involvement functions, while  Figure \ref{parameter_dis}(b) shows that cosine similarity outperforms Chebyshev similarity. Figure \ref{parameter_dis}(d) shows that Jaccard coefficient performs significantly better than average precision when \MEDOC~is considered, while \ Figure \ref{parameter_dis}(e) shows that the weighted association function is superior to the other for $\mu < 0.6$ and exhibits similar performance for $\mu\geq 0.6$.  We further vary the number of iterations $K$ to obtain communities with different vertex orderings -- Figures \ref{parameter_dis}(c) and \ref{parameter_dis}(f) show that for the networks with strong community structure (such as LFR ($\mu=0.1$), Football), the accuracy levels off at $K=0.2|V|$; however with increasing $\mu$ leveling off occurs at larger values of $K$. Note that the patterns observed here for the LFR network are similar for other networks. Therefore unless otherwise stated, in the rest of the experiment we show the results of our algorithms with the following parameter settings for disjoint community detection: \ENDISCO: $K=0.2|V|$, $RCC$, $COS$; \MEDOC: $K=0.2|V|$, $JC$, $F_w$.

\subsubsection{Impact of Base Community Detection Algorithms on \ENDISCO~and \MEDOC}\label{sec:impact}
In order to assess the  impact of each base algorithm in our ensemble, we measure the accuracy of \ENDISCO~and \MEDOC~when that base algorithm is removed from the ensemble in isolation --- Table \ref{impact} shows that for LFR and real-world networks \INF~has the biggest impact on accuracy according to both the evaluation measures (NMI and ARI) for both \ENDISCO~and \MEDOC. We also observe that the overall accuracy of the ensemble algorithms decrease after removing each base algorithm. Therefore, irrespective of the quality of the base algorithms, we might need all of them to obtain high quality community  structures as output.

However, from  this result it is not clear -- (i) whether we need all the outputs obtained from $K$ vertex orderings of a base algorithm, (ii) whether a certain combination of the base algorithms would produce the same accuracy as that obtained from using all the base algorithms. We will discuss more on these issues in Section \ref{base_selection}.

\begin{table}[!t]
 \centering

 \caption{Impact of each base algorithm on the accuracy of \ENDISCO~ and \MEDOC. The results are reported on default LFR and real networks with default parameter settings of the proposed algorithms (we use \INF~ as the final re-clustering algorithm). Each base algorithm is removed in isolation during the construction of ensemble matrix.}\label{impact}
 \scalebox{0.7}{
 \begin{tabular}{|l|l|c|c|c|c|c|c|}

\multicolumn{8}{c}{{\bf (a) LFR Network}} \\\hline
\multirow{3}{*}{No.} & Base & \multicolumn{4}{c|}{Disjoint} & \multicolumn{2}{c|}{Overlapping} \\\cline{3-8}
              &Algorithm & \multicolumn{2}{c|}{\ENDISCO} & \multicolumn{2}{c|}{\MEDOC} & \multicolumn{2}{c|}{\MEDOC} \\\cline{3-8}
            &  &  NMI & ARI & NMI & ARI & ONMI & $\Omega$ \\\hline
(1) & All & 0.85 & 0.89  & 0.87  & 0.90  & 0.84 & 0.87  \\
(2) & (1) $-$ \FG & 0.83 & 0.88 & 0.84 & 0.88 & 0.83 & 0.85 \\
(3) & (1) $-$ \LOU & 0.82 & 0.86 & 0.85 & 0.86 & 0.81 & 0.84 \\
(4) & (1) $-$ \CNM & 0.82 & 0.85 & 0.83 & 0.87 & 0.82 & 0.85 \\
\rowcolor[HTML]{D3D3D3}
(5) & (1) $-$ \INF & 0.80 & 0.81 & 0.81  & 0.82 & 0.80 & 0.81\\
(6) & (1) $-$ \WT & 0.84 & 0.88 & 0.85  & 0.81 & 0.83  & 0.86\\
(7) & (1) $-$ \LP & 0.84 &  0.87 & 0.86 & 0.87 & 0.83 & 0.85\\\hline

\multicolumn{8}{c}{{\bf (b) Real-world Network}} \\\hline
\multirow{3}{*}{No.} & Base & \multicolumn{4}{c|}{Football} & \multicolumn{2}{c|}{Senate} \\\cline{3-8}
              &Algorithm & \multicolumn{2}{c|}{\ENDISCO} & \multicolumn{2}{c|}{\MEDOC} & \multicolumn{2}{c|}{\MEDOC} \\\cline{3-8}
            &  &  NMI & ARI & NMI & ARI & ONMI & $\Omega$ \\\hline
(1) & All & 0.90 & 0.92  & 0.92  & 0.93  & 0.81 & 0.85  \\
(2) & (1) $-$ \FG & 0.89 & 0.90 & 0.91 & 0.90 & 0.80 & 0.83 \\
(3) & (1) $-$ \LOU & 0.87 & 0.86 & 0.88 & 0.91 & 0.78 & 0.80 \\
(4) & (1) $-$ \CNM & 0.86 & 0.87 & 0.88 & 0.91 & 0.79 & 0.82 \\
\rowcolor[HTML]{D3D3D3}
(5) & (1) $-$ \INF & 0.84 & 0.87 & 0.85  & 0.83 & 0.76 & 0.79\\
(6) & (1) $-$ \WT & 0.89 & 0.91 & 0.89  & 0.88 & 0.80  & 0.81\\
(7) & (1) $-$ \LP & 0.89 &  0.90 & 0.91 & 0.90 & 0.80 & 0.81\\\hline
 \end{tabular}}
\end{table}

\subsubsection{Impact of Re-clustering Algorithms on \ENDISCO~and \MEDOC}

As the final step in both \ENDISCO~and \MEDOC~is to run an  algorithm for re-clustering, we also conduct experiments to identify the best re-clustering algorithm.  Table~\ref{impact_den} shows that for both LFR networks and real networks, \INF~is the best re-clustering algorithm.

\begin{table}[!t]
 \centering

 \caption{Impact of each algorithm  at the final stage of \ENDISCO~ and \MEDOC~ to re-cluster vertices. The results are reported on (a) default LFR network and (b) two real-world networks with the default parameter values of the proposed algorithms.}\label{impact_den}
 \scalebox{0.7}{
 \begin{tabular}{|l|c|c|c|c|c|c|}
 
\multicolumn{7}{c}{{\bf (a) LFR Network}} \\\hline
Re-clustering & \multicolumn{4}{c|}{Disjoint} & \multicolumn{2}{c|}{Overlapping} \\\cline{2-7}
Algorithm & \multicolumn{2}{c|}{\ENDISCO} & \multicolumn{2}{c|}{\MEDOC} & \multicolumn{2}{c|}{\MEDOC} \\\cline{2-7}
            &  NMI & ARI & NMI & ARI & ONMI & $\Omega$ \\\hline

\FG  &  0.79 & 0.80 & 0.80 & 0.83 & 0.81 & 0.84\\
\LOU & 0.82 & 0.84 & 0.83 & 0.86 & 0.82 & 0.83\\
\CNM & 0.83 & 0.81 & 0.83 & 0.86 & 0.81 & 0.80\\
\rowcolor[HTML]{D3D3D3}
\INF & 0.85 & 0.89  & 0.87  & 0.90  & 0.84 & 0.87  \\
\WT & 0.75 & 0.78 & 0.77 & 0.82 & 0.76 & 0.79\\
\LP & 0.77 & 0.79 & 0.78 & 0.80 & 0.75 & 0.77\\\hline
\multicolumn{7}{c}{{\bf (b) Real-world Network}} \\\hline
Re-clustering & \multicolumn{4}{c|}{Football} & \multicolumn{2}{c|}{Senate} \\\cline{2-7}
Algorithm & \multicolumn{2}{c|}{\ENDISCO} & \multicolumn{2}{c|}{\MEDOC} & \multicolumn{2}{c|}{\MEDOC} \\\cline{2-7}
            &  NMI & ARI & NMI & ARI & ONMI & $\Omega$ \\\hline

\FG  &  0.86 & 0.89 & 0.87 & 0.91 & 0.78 & 0.77\\
\LOU & 0.87 & 0.91 & 0.88 & 0.89 & 0.79 & 0.84\\
\CNM & 0.87 & 0.90 & 0.89 & 0.90 & 0.80 & 0.81\\
\rowcolor[HTML]{D3D3D3}
\INF & 0.90 & 0.92  & 0.92  & 0.93  & 0.81 & 0.85  \\
\WT & 0.81 & 0.87 & 0.84 & 0.82 & 0.76 & 0.79\\
\LP & 0.78 & 0.79 & 0.80 & 0.82 & 0.75 & 0.78\\\hline
 \end{tabular}}
\end{table}

\subsubsection{Comparative Evaluation}\label{comp_dis}

Table \ref{performance:non_synthe} and Table \ref{performance:non_real} report the performance of our approaches on synthetic and real-world networks respectively using different algorithms in the final step of \ENDISCO, \MEDOC~ and \CC~  for synthetic networks. The numbers denote relative performance improvement of \ENDISCO, \MEDOC~ and \CC~ with respect to
a given algorithm when that algorithm is used in the final step. For instance, the last entry in the last row (7.82) means that for LFR ($\mu=0.6$) network, the accuracy of \MEDOC~(when \LP\ is used for re-clustering in its final step) averaged over NMI and ARI is 7.82\% higher than that of the standalone \LP. The actual values are reported in Table \ref{actual_result}. The point to take away from this table is that irrespective of which classical community detection algorithm we compare against, \ENDISCO~and \MEDOC~always improve the quality of communities found. Moreover, our proposed algorithms outperform \CC~ for all the networks. We further observe from the results of LFR networks that with the deterioration of the community structure (increase of $\mu$), the improvement increases for all the re-clustering algorithms. This essentially indicates that ensemble based approaches are even more useful if the underlying community structure is not well-separated. 

We further compare \ENDISCO\ and \MEDOC\ with the performance of competing algorithms on real-world networks. Table \ref{performance:non_real} presents the same patterns observed for synthetic networks. Once again, our proposed algorithms outperform both \CC~ and the other standalone algorithms. For the Football network, our algorithms perform as well as other baseline algorithms, because the underlying community structure is very clear in the Football network \cite{Clauset2004}. Interestingly, our algorithms exhibit better performance for the Coauthorship network which has weaker community structure \cite{Chakraborty:2014}. 

Both the results on synthetic and real-world networks lead to the conclusions that -- (i) \ENDISCO\ and \MEDOC\ algorithms are quite competitive to the standalone algorithms for those networks which have prominent community structure; (ii)  \ENDISCO\ and \MEDOC\ algorithms are more effective than the standalone algorithms for those networks whose underlying communities are weakly separated and difficult to detect by any traditional community detection algorithm.

\begin{table}[!t]
\centering
\caption{Relative percentage improvement (averaged over NMI and ARI) of \CC, \ENDISCO~ and \MEDOC~ over the baseline algorithms for disjoint community detection from synthetic networks. Each row corresponds to an algorithm $Al$ and the value indicates the performance improvement of the ensemble approach with $Al$ as the re-clustering algorithm over the isolated performance of $Al$ without ensemble.}  \label{performance:non_synthe}
\scalebox{0.7}{
\begin{tabular}{|c|cc>{\columncolor[gray]{0.8}}c|cc>{\columncolor[gray]{0.8}}c|cc>{\columncolor[gray]{0.8}}c|}
\hline
 \multirow{3}{*}{Algorithm} & \multicolumn{9}{c|}{Synthetic Network}\\\cline{2-10}
  & \multicolumn{3}{c|}{LFR ($\mu=0.1$)} & \multicolumn{3}{c|}{LFR ($\mu=0.3$)} & \multicolumn{3}{c|}{LFR ($\mu=0.6$)} \\\cline{2-10}
  & \CC & \ENDISCO & \MEDOC & \CC & \ENDISCO & \MEDOC & \CC & \ENDISCO & \MEDOC\\\hline
\FG & 1.92 & 2.39 & 2.93 & 1.96 & 2.71 & 3.02 & 1.90 & 3.81 & 3.91 \\   
\LOU & 1.86 & 1.97 & 2.04 & 1.90 &  2.22  & 2.40 & 1.97 &  3.41 & 3.86\\
\CNM  & 1.98 & 2.07  & 2.46 & 2.03 & 2.14 & 2.83 & 2.01 &3.22 &    3.50\\
\INF  &0 & 0 & 0 & 0.98  & 1.44 & 1.62 & 1.62 &  2.01 & 2.46\\
\WT   &3.43 &4.43 & 4.97 & 3.91  & 4.86 & 5.08 & 5.05& 6.98 & 7.42 \\ 
\LP   &3.90 & 5.06 & 5.72 & 4.01 & 5.12 & 5.39  & 4.96 & 7.50 & 7.82\\\hline  
\end{tabular}}

\end{table}

\begin{table}[!t]
\centering
\caption{Relative percentage improvement (averaged over NMI and ARI) of \CC, \ENDISCO~ and \MEDOC~ over the baseline algorithms for disjoint community detection from real-world networks. Each row corresponds to an algorithm $Al$ and the value indicates the performance improvement of the ensemble approach with $Al$ as the re-clustering algorithm over the isolated performance of $Al$ without ensemble.}  \label{performance:non_real}
\scalebox{0.6}{
\begin{tabular}{|c|cc>{\columncolor[gray]{0.8}}c|cc>{\columncolor[gray]{0.8}}c|cc>{\columncolor[gray]{0.8}}c|cc>{\columncolor[gray]{0.8}}c|}
\hline
 \multirow{3}{*}{Algorithm} & \multicolumn{12}{c|}{Real-world Network}\\\cline{2-13}
  & \multicolumn{3}{c|}{Football} & \multicolumn{3}{c|}{Railway} & \multicolumn{3}{c|}{University} & \multicolumn{3}{c|}{Coauthorship} \\\cline{2-13}
  & \CC & \ENDISCO & \MEDOC & \CC & \ENDISCO & \MEDOC & \CC & \ENDISCO & \MEDOC & \CC & \ENDISCO & \MEDOC\\\hline
\FG & 0 &  0 &  0   &  1.01& 1.22 &   1.43   & 1.92& 2.20 & 2.86 & 2.23  & 3.98  & 4.60\\
\LOU & 0 & 0 & 0 & 0.96 & 1.17 & 1.43 & 1.76 & 2.12 & 2.30 & 1.87  & 2.21 & 2.39\\
\CNM & 0.84 & 1.23 & 1.46  & 1.01 & 1.49 & 1.92 & 1.54 & 2.39 &  2.40 & 1.32  & 2.92 & 3.41\\
\INF & 0 & 0 & 0 & 1.10 & 1.22 & 1.56 & 1.76 & 2.01 & 2.20 & 1.98 & 2.31 & 2.98\\
\WT  & 1.42 & 2.21 & 2.46 & 2.13 & 3.21 & 3.49 & 3.01 &   4.22 &   4.49 & 4.02 & 5.06 & 5.51\\
\LP  & 2.23 & 3.01 &  3.29 & 2.21 &  3.46 & 3.79 & 4.32 & 6.21 & 6.80 & 4.32  & 6.21          & 6.98\\\hline
\end{tabular}}
\end{table}


\begin{table}[!t]
 \centering

 \caption{Actual accuracy values (in terms of NMI and ARI) of the proposed ensemble approaches (with default parameter setting) and \CC~  on both synthetic and real-world networks.}\label{actual_result}
 \scalebox{0.6}{
  \begin{tabular}{|l|c|c|c|c|c|c|}
 \hline

Network     & \multicolumn{2}{c|}{\CC} & \multicolumn{2}{c|}{\ENDISCO} & \multicolumn{2}{c|}{\MEDOC} \\\cline{2-7}
            & NMI & ARI &  NMI & ARI & NMI & ARI \\\hline
LFR($\mu=0.1$) & 0.89 & 0.91 & 0.93  & 0.96   &  0.94  &  0.96 \\
LFR($\mu=0.3$) & 0.85 & 0.86 & 0.89  & 0.88  & 0.90   & 0.91\\
LFR($\mu=0.6$) & 0.76 & 0.79 & 0.82  & 0.84  & 0.84   &  0.86\\\hline
Football       & 0.90 & 0.92 & 0.90 & 0.92 & 0.92 & 0.93 \\
Railway        & 0.71 & 0.73 & 0.78& 0.80 &0.79 & 0.83 \\
University     & 0.75 & 0.79 & 0.83 & 0.86 & 0.86 & 0.87\\
Coauthorship   & 0.61 & 0.65  & 0.67 & 0.68 & 0.70& 0.76 \\\hline            
 \end{tabular}}
\end{table}

\section{Experiments: Results for Overlapping Community Detection}\label{sec:overlapping_result}
In this section, we evaluate \MEDOC~for overlapping community detection. We start by explaining the datasets used in this experiment, followed by the baseline algorithms used to compare our method and the evaluation metrics used to compare the detected communities with the ground-truth. We then show how to choose the best parameters followed by the comparative evaluation.

\subsection{Datasets}
Here we briefly describe the synthetic and real-world networks that we use in this experiment.
\subsubsection{Synthetic Networks}
We again use the  LFR benchmark to generate synthetic networks with overlapping community structure with the following default parameter settings as mentioned in \cite{oslom,Gregory1}: $n=10000$, $\bar k=50$, $k_{max}=150$, $\mu=0.3$, $O_n=20\%$, $O_m=20$, $c_{max}=100$, $c_{min}=20$.
We generate $50$ LFR networks for each parameter configuration --- the experiments reported averages over these $50$ networks. We further vary $\mu$ ($0.1$-$0.8$ with increment of $0.05$), $O_m$ and $O_n$ (both from $15\%$ to $30\%$ with increment of $1\%$) depending upon the  experimental need.

\begin{table}
\caption{Properties of the real-world networks with overlapping community structure. $N$: number of vertices, $E$: number of edges, $C$: number of
communities, $\rho$: average edge-density per community, $S$: average size of a community, $O_m$: average number of community memberships per vertex.}\label{dataset_over}
\centering
\scalebox{0.85}{
\begin{tabular}{l||r|r|r|r|r|r}
\hline
Network & $N$ & $E$ & $C$ & $\rho$ & S & $O_m$ \\\hline
Senate & 1,884 & 16,662 & 110    & 0.45 & 81.59 & 4.76 \\ 
Flickr  & 80,513 & 5,899,882 & 171 & 0.046 & 470.83 & 18.96  \\
Coauthorship  &  391,526 & 873,775 & 8,493 & 0.231 & 393.18& 10.45 \\
LiveJournal  & 3,997,962 & 34,681,189 & 310,092 & 0.536  & 40.02 & 3.09  \\
Orkut   & 3,072,441 & 117,185,083 & 6,288,363 & 0.245 & 34.86 & 95.93 \\\hline
\end{tabular}}
\end{table}

\subsubsection{Real-world Networks}

We also run experiments with following six real-world datasets mentioned in Table \ref{dataset_over}:

\textbf{Senate network:} This network combines voting pattern of $110$ US-Senates \cite{PhysRevE.91.012821,KlosterG15}. Each vertex represents a senator and the senators are connected in that session to their $3$ nearest neighbors measured by voting similarities. The ground-truth communities are marked based on the senators who served in the same instance of the senate, i.e., senators who served during the same term.

\textbf{Flickr:} This dataset is built by forming links between images sharing common metadata from Flickr \cite{Wang-etal12}. Edges are formed between images from the same location, submitted to the same gallery, group, or set, images sharing common tags, images taken by friends, etc. Communities are the user-specified groups.

\textbf{Coauthorship:} The coauthorship network \cite{Palla} here is exactly the same as mentioned before for disjoint community structure in Section \ref{coau} except the ground-truth community marking which in this case is the publication venues (conferences or journals).

\textbf{LiveJournal:} LiveJournal is a free on-line blogging community where users declare their friends. LiveJournal also allows users to form a group which other members can then join. Here user-defined groups are considered as ground-truth communities.  \cite{Leskovec} provided the LiveJournal friendship social network and ground-truth communities.

\textbf{Orkut:}
In this datasets, users in Orkut social networking site are nodes, and links are their friendships. Ground-truth communities are user-defined groups. \cite{Leskovec} provided the Orkut friendship social network and ground-truth communities.

\subsection{Baseline Algorithms}
There are several standalone overlapping community detection algorithms, which are different based on the underlying working principle. We take six  state-of-the-art algorithms from three different categories mentioned in \cite{Xie:2013}: (i)  {\bf   Local expansion:} OSLOM \cite{oslom} and EAGLE \cite{Shen}; (ii)  {\bf Agent-based dynamical algorithms:}  COPRA \cite{Gregory1} and SLPA \cite{Xie}, (iii)  {\bf Detection using mixture model:}  MOSES \cite{moses} and BIGCLAM \cite{Leskovec}.

\subsection{Evaluation Metrics}\label{metric_overlapping}
To compare the detected overlapping community structure with the ground-truth, we consider two standard validation metrics: Overlapping Normalized Mutual Information (ONMI)\footnote{\url{https://github.com/aaronmcdaid/Overlapping-NMI}} cite{Lancichinetti,journals} and  Omega Index ($\Omega$ Index) \cite{collins1988,Murray:2012}. The larger the value of ONMI and Omega index, the better the matching between two community structures.

\begin{figure}[!t]
\centering
 \scalebox{0.3}{
  \includegraphics{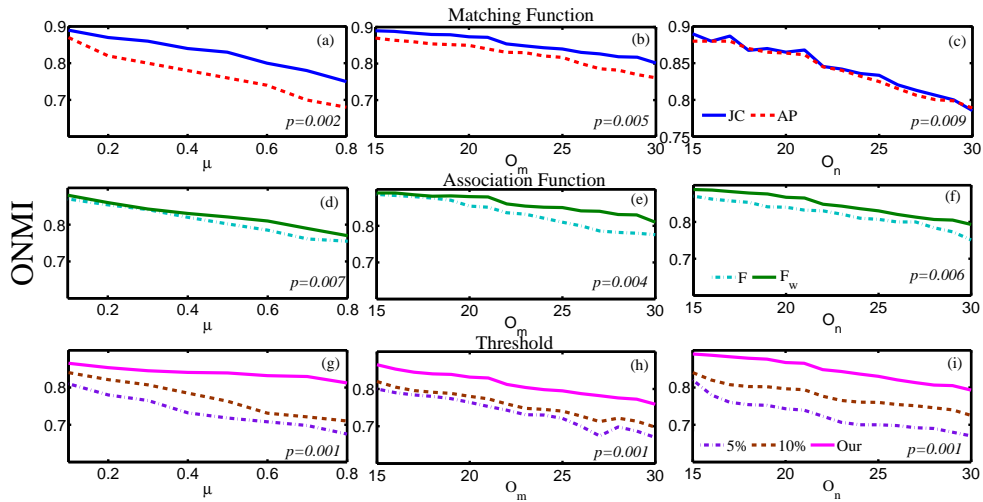}}
  \caption{Dependencies of \MEDOC~ on different algorithmic parameters. The results are reported on default overlapping LFR networks by varying three parameters $\mu$, $O_m$ and $O_n$. For thresholding, we choose top 5\% and 10\% highly probable communities for each vertex and compare it with our threshold selection method.  The value corresponding to one parameter is reported by averaging the values for all possible combinations of the other parameters. The results are statistically significant.}\label{parameter_over}
\end{figure}

\subsection{Experimental Results}
In this section,  we present a detailed description of the comparative evaluation of the competing algorithms for overlapping community detection. We first describe the parameter selection process for \MEDOC, followed by the results showing the impact of the base algorithms on \MEDOC. We then  present the performance of the competing algorithms.

\subsubsection{Parameter Settings}
We first try to identify the best parameter settings for \MEDOC.  These include: 
matching function $W$, association function $\mathcal{F}$,  number of iterations $K$ and threshold $\tau$. Figure \ref{parameter_over} shows the results on LFR networks by varying $\mu$, $O_m$ and $O_n$. The results are almost identical with that of disjoint setting shown in Figure \ref{parameter_dis}. 
We observe that Jaccard coefficient as matching function and weighted association measure are better than their respective alternatives. The choice of $K$ is the same as shown in Figure \ref{parameter_dis}(f) -- accuracy almost levels off at $K=0.2|V|$. We experiment with two choices of thresholding: top 5\% and 10\% most probable communities per vertex, and compare with the threshold selection mechanism described in Section \ref{algo2:Parameter}. Figures \ref{parameter_over}(g) and \ref{parameter_over}(i) show that irrespective of any network parameter selection, our choice of selecting threshold always outperforms others. As shown in Table \ref{impact_den}, \INF~ seems to be the best choice for the re-clustering algorithm. Therefore unless otherwise stated, in the rest of the experiments, we  run \MEDOC~with $K=0.2|V|$, $JC$, $F_w$, \INF~ and $\tau$ (selected by our method).

\subsubsection{Impact of Base Algorithms for Overlapping Community Detection}
The impact of the base algorithms on \MEDOC's performance is similar to what we saw in the disjoint community detection case. The results in Table \ref{impact} show that accuracy decreases most when we drop \INF~from the base algorithm, followed by \LOU~ and \CNM.


\begin{table}[!t]
\begin{center}
\caption{Accuracy of all the competing algorithms in detecting the overlapping community structure from synthetic networks. All the disjoint algorithms are used to create  the multipartite network and \MEDOC~ is run with its default parameter setting. } \label{performance:over_synthetic}
\scalebox{0.7}{
\begin{tabular}{|l|cccccccc|}
\hline

\multirow{2}{*}{Algorithm} & \multicolumn{8}{c|}{{\bf Synthetic Networks}} \\
\multirow{2}{*}{ } & \multicolumn{2}{c}{LFR ($\mu=0.1$)} & &\multicolumn{2}{c}{LFR ($\mu=0.3$)} & &\multicolumn{2}{c|}{LFR ($\mu=0.6$)} \\ \cline{2-3} \cline{5-6} \cline{8-9}  
                           & ONMI               & $\Omega$     &         & ONMI               & $\Omega$ &              & ONMI               & $\Omega$\\ \cline{1-9}

\OSLOM                 & 0.80	&0.78&&	0.74&	0.78&&	0.72&	0.73  \\                           

\EAGLE     &  0.81	&0.83	&&0.75&	0.76&&	0.70	&0.74\\
\COPRA    & 0.80	&0.81&&	0.76&	0.74&&	0.72&	0.74\\

\SLPA       & 0.84&	0.86&&	0.78&	0.77&&	0.76&	0.77\\

\MOSES            & 0.85&	0.86	&&0.80&	0.81&&	0.75&	0.78 \\

\BIGCLAM         & 0.86	&0.85	&&0.81&	0.83&&	0.77&	0.79  \\
\MEDOC         &  \cellcolor{gray!25} 0.88&	\cellcolor{gray!25}0.91	&&\cellcolor{gray!25}0.84&	\cellcolor{gray!25}0.87&&	\cellcolor{gray!25}0.82&	\cellcolor{gray!25}0.84\\

\hline

\hline
\end{tabular}}
 
\end{center}
\end{table}

\begin{table}[!t]
\begin{center}
\caption{Accuracy of all the competing algorithms in detecting the overlapping community structure from real-world networks. All the disjoint algorithms are used to create  the multipartite network and \MEDOC~ is run with its default parameter setting. } \label{performance:over_real}
\scalebox{0.7}{
\begin{tabular}{|l|cccccccccccccc|}
\hline
\multirow{2}{*}{Algorithm} & \multicolumn{14}{c|}{{\bf Real-world Networks}} \\

\multirow{2}{*}{ }  &\multicolumn{2}{c}{Senate} & &\multicolumn{2}{c}{Flickr} & &\multicolumn{2}{c}{Coauthorship} & &\multicolumn{2}{c}{LiveJournal} & &\multicolumn{2}{c|}{Orkut} \\ \cline{2-3} \cline{5-6} \cline{8-9} \cline{11-12} \cline{14-15} 

                           & ONMI           & $\Omega$&           & ONMI           & $\Omega$&          & ONMI            & $\Omega$&            & ONMI             & $\Omega$&     & ONMI & $\Omega$  \\ \hline

\OSLOM                 &	0.71&	0.73	&&0.68&	0.74&&	0.70&	0.71&&	0.73&	0.75&&	0.71	&0.76

      \\

\EAGLE     &	0.73	&0.74&&	0.69	&0.76&&	0.71&	0.74&&	0.74&	0.76&&	0.70	&0.77
        \\
\COPRA    &	0.74	&0.77&&	0.73&	0.78&&	0.75&	0.79&&	0.76&	0.82&&	0.74&	0.76
  \\

\SLPA      &	0.74&	0.76&&	0.72&	0.74&&	0.76&	0.77&&	0.78&	0.85&&	0.75&	0.79\\

\MOSES          &	0.75&	0.78&&	0.74&	0.76&&	0.79&	0.78&&	0.81&	0.82&&	0.78&	0.82
   \\

\BIGCLAM         &	0.76&	0.79&&	0.75&	0.76&&	0.80&	0.84&&	0.84&	0.87&&	0.81&	0.84
     \\
\MEDOC         &	\cellcolor{gray!25}0.81&	\cellcolor{gray!25}0.85&&	\cellcolor{gray!25}0.79&	\cellcolor{gray!25}0.84&&	\cellcolor{gray!25}0.82&	\cellcolor{gray!25}0.86&&	\cellcolor{gray!25}0.86&	\cellcolor{gray!25}0.88&&	\cellcolor{gray!25}0.83&	\cellcolor{gray!25}0.86
  \\

\hline

\hline
\end{tabular}}
 
\end{center}
\end{table}

\subsubsection{Comparative Evaluation}
We run \MEDOC~with the default setting on three LFR networks and five real-world networks. The performance of \MEDOC~is compared with the six baseline overlapping community detection algorithms. Table \ref{performance:over_synthetic} shows the performance of the competing algorithms in terms of ONMI and $\Omega$ index for synthetic network. In all cases, \MEDOC~is a clear winner, winning by significant margins. The absolute average of ONMI ($\Omega$) for \MEDOC~ over all synthetic networks is 0.85	(0.87), which is 4.50\% (5.66\%) higher than \BIGCLAM,  6.25\% (6.53\%) higher than \MOSES, 7.14\% (8.75\%) higher than \SLPA, 11.84\% (13.97\%) higher than \COPRA, 12.83\% (12.01\%) higher than \EAGLE, and 12.38\% (13.97\%) higher than \OSLOM. Another interesting observation is that for synthetic networks, the more the community structure deteriorates with the increase in $\mu$, the harder it becomes to detect the communities. It is in then ``hard to detect'' cases that  the performance of \MEDOC~significantly improves compared to the baselines. Another interesting observation is that the performance improvement seems to be prominent with the deterioration of community quality. For instance, the improvement of \MEDOC~with respect to the best baseline algorithm (\BIGCLAM) is 2.32\% (7.06\%), 3.70\% (4.82\%) and 6.49\% (6.33\%) in terms of ONMI ($\Omega$) with the increasing value of $\mu$ ranging from 0.1, 0.3 and 0.6 respectively. This once again corroborates our earlier observations in Section \ref{comp_dis} that \MEDOC~is highly effective for those networks where the underlying community structure is not prominent and hard to detect.

In Table \ref{performance:over_real}, we show the performance of the competing algorithms on real-world networks. \MEDOC~once again outperforms other competing methods. The average absolute ONMI of \MEDOC~over all networks is 0.82, which is followed by 
\BIGCLAM~ (0.79), \MOSES~(0.77), \OSLOM~(0.76), \SLPA~(0.75), \COPRA~(0.74) and \EAGLE~(0.71). In short, \MEDOC~performs the best irrespective of any network and used validation measure.

\section{Experiments: Results of Fuzzy Community Detection}\label{sec:fuzzy_result}
In the case of overlapping communities, there are two different ways of defining overlap -- in {\em crisp overlapping} in which each vertex belongs to one or more communities (the membership is binary -- there is no notion of the strength of membership in a communtiy); whereas in case of {\em fuzzy overlapping}, a vertex  may also belong to more than one community but the strength of its membership to each community can vary.   
For instance, a person on Facebook might belong to multiple groups, but he may be much more active in one group compared to another. In this case, his degree of  membership in that one group would be considered to be larger.

We earlier observed in Step \ref{fuzzy} of \MEDOC~that it can assign each vertex $v$ to a community $C$ with a membership probability of $\hat{\mathbb{A}} (v,C)$. In this section, we provide a comprehensive analysis of the performance of \MEDOC~in detecting fuzzy community structure. We start by explaining the datasets used in this experiment, followed by the baseline algorithms and the evaluation metrics. We then describe the results of our comparative experimental evaluation.

\subsection{Datasets}
We first describe the construction of synthetic networks with fuzzy community structure, followed by the real-world network.

\subsubsection{Synthetic Network} The synthetic network generated by the LFR model \cite{PhysRevE} does not contain the fuzzy community structure. \cite{Gregory2011} proposed a modified version of the LFR model to generate synthetic fuzzy community structure. Here we adopt the their approach to generate synthetic networks. First, we generate crisp overlapping communities from the LFR model. Second, the crisp communities are converted to fuzzy form by adding a random membership probability to each occurrence of a vertex. The membership probabilities are chosen from a uniform distribution. Next a network is constructed from the fuzzy communities
using the following formula: 
\begin{equation}
 p_{ij}=s_{ij}p_1+(1-s_{ij})p_0
\end{equation}
where $p_{ij}$ is the probability of the existence of an edge $e_{ij}$, $s_{ij}$ is the co-membership of vertices $i$ and $j$; and $p_{ij}=p_1\ if\ \exists c\in C [i\in c \wedge j \in c];\ else\ p_0$. In the above equation, $p_0$ and $p_1$ are chosen so as to preserve the specified average degree ($<k>$) and mixing parameter ($\mu$) in the generated LFR network. The final network then satisfies all of the original parameters of LFR with the exception of the degree distribution ($k_{max}$), maximum degree and $\tau_1$, the exponent of the power-law distribution of vertex degrees). However, other parameters of LFR (such as $\mu$, $O_m$, $O_n$ etc.) represent the same functionalities in this model. More details can be found in \cite{Gregory2011}\footnote{We took the implementation of the synthetic model by the author available at \url{http://www.cs.bris.ac.uk/~steve/networks/}.}. Unless otherwise stated, we generate the synthetic networks with the following parameter setting: $n=10000$, $\bar k=50$, $k_{max}=150$, $\mu=0.3$, $O_n=20\%$, $O_m=20$, $c_{max}=100$, $c_{min}=20$.

\subsubsection{Real-world Network} There is no real-world network where fuzzy community memberships of vertices are known. Therefore, the existing fuzzy community detection algorithms were mostly tested either with synthetic networks \cite{Gregory2011,TVCG.2013.232}, or by calculating community evaluation metrics such as modularity \cite{6891611}. Here we use the metadata information of the coauthorship network mentioned in Section \ref{coau} to construct the ground-truth. We recall that in the coauthorship network, authors are the vertices, edges are drawn based on coauthorship relations, and communities are different research areas. We then assign each author into a community with the community membership indicated by the fraction of papers the author has written on the corresponding research area. It also ensures that the sum of community memberships of each author is $1$.

\subsection{Baseline Algorithms}

Fewer fuzzy methods have been proposed in the past. \cite{14} presented ``FuzzyClust'' that maps the problem to a nonlinear constrained optimization problem and solves it. \cite{11}  used the fuzzy $c$-means algorithm to detect up to communities after converting the network into the feature space. \cite{16} presented a method based on Bayesian non-negative matrix factorization (NMF). Finally, FOG \cite{DavisC08} clusters ``link data'', which includes networks as a special case, into fuzzy communities based on stochastic framework. 
\cite{Gregory2011} suggested ``MakeFuzzy'' algorithm  which is used as a post-processing technique after a crisp overlapping algorithm to detect the fuzzy community structure. He further showed that MakeFuzzy along with \EAGLE~ \cite{Shen} outperforms other state-of-the-art algorithms.

In our experiment, we consider FuzzyClust algorithm of \cite{14} and the NMF algorithm of \cite{16} (with default parameters) as two baseline algorithms. We also consider MakFuzzy+\EAGLE ~(henceforth, named as ``\MAKGLE'') as another baseline to compare  with \MEDOC.

\subsection{Evaluation Metric}
There exist very few metrics for comparing two fuzzy community structures. As per as we are aware, Fuzzy Rand Index (FRI) proposed by \cite{HullermeierRHS12} is the only one metric for this purpose. Since this metric is used infrequently, we here explain this metric. This is a redefined version of the original Rand Index \cite{Gregory2011}:
\begin{equation}
 RI_u(C_1,C_2)=\frac{s(C_1,C_2)}{N}
\end{equation}
where $s(C_1,C_2)=N-\sum_{i,j\in V} |f(i,j,C_1)-f(i,j,C_2)|$,  and $f(i,j,C_1)=1$ is vertices $v_i$ and $v_j$ appear in the same community, 0 otherwise. Then the expected Rand Index is defined as: $RI_e(C_1,C_2)=\frac{s(C_1)s(C_2)+(N-s(C_1))(N-s(C_2))}{N^2}$, where $N$ is the total number of vertices and  $s(C)=\sum_{i,j\in V}f(i,j,C)$.

The function $f(i,j,C)$ indicates the extent to which $i$ and $j$ appear in the same community in $C$, which depends on the membership probability of $v_i$ and $v_j$ as follows:

\begin{equation}
 f(i,j,C)=1-\frac{1}{2} \sum_{c\in C}[\alpha_{ic}-\alpha_{jc}]
\end{equation}
where $\alpha_{ic}$ is the membership probability of $i$ in community $c$.


\begin{table}[!t]
 \centering
 \caption{Accuracy (in terms of Fuzzy Rand Index) of the fuzzy community detection algorithms for both synthetic and real-world networks. Synthetic networks are generated by varying $\mu$ and setting other parameters to default values. We run \MEDOC~in two settings -- (i) \MEDOC: Algorithm \ref{meclud} to detect fuzzy communities, (ii) \MEDOC +MakeFuzzy: crisp overlapping communities are detected by \MEDOC, followed by \MAKGLE~to post-process the output.}\label{result_fuzzy} 
 \scalebox{0.8}{
 \begin{tabular}{|l|ccc|c|}
 \hline
  \multirow{2}{*}{Algorithm} & \multicolumn{3}{c}{Synthetic} & \multicolumn{1}{|c|}{Real-world}\\\cline{2-5}
     & LFR ($\mu=0.1$) & LFR ($\mu=0.3$) & LFR ($\mu=0.6$) & Coauthorship \\\hline
 FuzzyClust  & 0.71   & 0.68   &  0.65  & 0.62  \\
 NMF   & 0.74      & 0.70   & 0.68   & 0.65\\
 \MAKGLE  & {\bf 0.78} & 0.74 & 0.71 & 0.67\\
 \MEDOC & {\bf 0.78} & 0.75  & 0.70 & {\bf 0.68} \\
 \MEDOC + MakeFuzzy & {\bf 0.78} & {\bf 0.76} & {\bf 0.73} & {\bf 0.68}\\ \hline
 \end{tabular}}
\end{table}

\subsection{Experimental Results}
We choose the same parameters for \MEDOC~as shown in Figure \ref{parameter_over}, i.e., $K=0.2|V|$, $JC$ as pair-wise similarly of communities, $F_w$ as weighted association measure, \INF~as re-clustering algorithm. Further, we consider two setups for \MEDOC: (i) \MEDOC~ is run with the default parameter setting to detect the fuzzy community structure, (ii) \MEDOC~ is run to detect the crisp overlapping community first, and then MakeFuzzy is used as a post-processing technique to detect the fuzzing overlapping communities (we call it \MEDOC + MakeFuzzy). 

Table \ref{result_fuzzy} presents the results of three baseline algorithms along with two setups of \MEDOC. We observe that both the setups of \MEDOC~ tend to be very competitive with the \MAKGLE, which seems to be the best baseline algorithm. However, incorporating MakeFuzzy into \MEDOC~outperforms other competing algorithms with a significant margin.

\section{Selection of Base Outputs}\label{base_selection}
In Section \ref{sec:impact}, we observed that removal of each base algorithm from the entire set reduces the overall accuracy of the ensemble algorithms with a certain extent. However, it was not clear (i) whether we need to consider {\em all $K$ outputs} obtained from running each base algorithm $K$ times, (ii) whether a subset of base algorithms are enough to get similar accuracy. In this section, we address these questions. In particular, we ask a general question - {\em given a large set of different base solutions, how do we select a subset of solutions to form a smaller yet better performing ensemble algorithm than using all the available base solutions?}

To this end, we investigate two properties of the detected solutions that have already been identified to be effective in literature \cite{Parsons:2004,1399790,Hadjitodorov:2006,SAM:SAM10008} -- the {\em quality} and the {\em diversity}.

\subsection{Defining Quality and Diversity}
{\bf Quality.} Since the original communities to which vertices in a network belong are not known a priori, we propose to use an internal quality measure as follows. Given an ensemble solution $E$ combining the set of all base community structures $\Gamma=\{\mathbb{C}_m^k\}, \forall m\in M \wedge k\in K$, the following quality function is used to measure the similarity of each solution with the ensemble:
\begin{equation}
 Quality(\mathbb{C}_m^k,\Gamma)=\sum_{m'\in M} \sum_{n=1}^K \mathbb{Q}(\mathbb{C}_m^k,\mathbb{C}_{m'}^n) 
\end{equation}

$\mathbb{Q}$ can be any of the standard evaluation metrics such as NMI, ARI mentioned in Section \ref{dis_eval} for disjoint communities and ONMI, Omega index in Section \ref{metric_overlapping} for overlapping communities. However, we use NMI and ONMI for disjoint and overlapping communities respectively as suggested by Strehl and Ghosh \cite{Strehl:2003}.
Intuitively, $Quality$ measures how well a particular base solution agrees with the general trend present in $\Gamma$.\\

\noindent{\bf Diversity.} There are many different diversity measures proposed for cluster ensembles in data mining \cite{Li2010}. However, to make the function consistent with the quality measure, we consider pair-wise NMI/ONMI among the base solutions. In particular, we measure the pair-wise similarity of two base solutions as $\mathbb{Q}(\mathbb{C}_m^k, \mathbb{C}_{m'}^{k'})$ and compute the sum of all pair-wise similarities $\sum_{i\neq j \wedge i,j\in M \wedge k,k'\in K \wedge \mathbb{C}_i^k, \mathbb{C}_j^{k'}\in \Gamma} \mathbb{Q}(\mathbb{C}_i^k, \mathbb{C}_j^{k'})$. The lower the value, the higher the diversity. We consider this diversity measure because it has already been shown to be effective for cluster ensemble \cite{FernB03a}. 

Note that the base solution selection strategies that we will present here do not limit themselves to any particular quality and diversity functions.

\subsection{Selection Strategies}
Among the $MK$ number of solutions obtained from base community detection algorithms, we select $S$ solutions based on the following criteria individually:

\subsubsection{Only Quality} This strategy simply ranks all solutions based on $Quality$ and selects top $S$ solutions to include in the ensemble algorithm. The solution with the highest $Quality$ essentially indicates that it has high consistency with the rest of the solutions. Generally, we expect that if we  take only high quality solutions, due to the high similarity among them this strategy may lead to huge redundancy in the chosen set of solutions. This in turn reduces the ability to obtain improved results for those portions of the data which none of the selected solutions is able to capture properly. This explains the need for diversity amongst the solutions.

\subsubsection{Only Diversity} We consider a greedy solution to select $S$ solutions which are highly diverse. We start by selecting the solution with highest $Quality$\footnote{However, one can start by selecting the solution which has highest pair-wise diversity. However, we observed that it ended up selecting poor solutions in the ensemble which leads to decrease in performance.}. We then incrementally add one solution at a time such that the resulting ensemble has the highest diversity. This process continues until the required number of solutions are chosen. It is commonly believed that diversifying the base solution is beneficial because mistakes made by one base algorithm may be compensated by another. However, it may result in the inclusion of low quality solutions in the ensemble. This is one of the reasons to choose the solution with highest quality first in the greedy strategy.

\subsubsection{Combining Quality and Diversity} There is always a trade-off between quality and diversity in selecting base solutions, which can be viewed as a multi-objective optimization framework. A traditional way of handling such problems is to combine them into a single aggregate objective function. We choose $S$ such solutions, denoted by $\mathbb{C}_S$ that maximize the following objective function:
\begin{equation}
 J=\underbrace{\alpha \sum_{c\in \mathbb{C}_S} Quality(c,\Gamma)}_\text{Quality} + \underbrace{(1-\alpha)\sum_{c_i,c_j\in \mathbb{C}_S, c_i\neq c_j, } (1-\mathbb{Q}(c_i,c_j))}_\text{Diversity}
\end{equation}

The parameter $\alpha$ controls the trade-off between these two quantities. However, it is computationally expensive to select $S$ solutions out of $MK$ solutions  \cite{FernL08}.
 Therefore, we adopt the following greedy strategy. We start by adding the solution with highest quality and incrementally add solutions one at a time that maximizes $J$. We set $0.5$ as the default value of $\alpha$. However, we will examine different choices of $\alpha$ in Figure \ref{alpha}.

\begin{figure}
 \centering
 \includegraphics[width=\columnwidth]{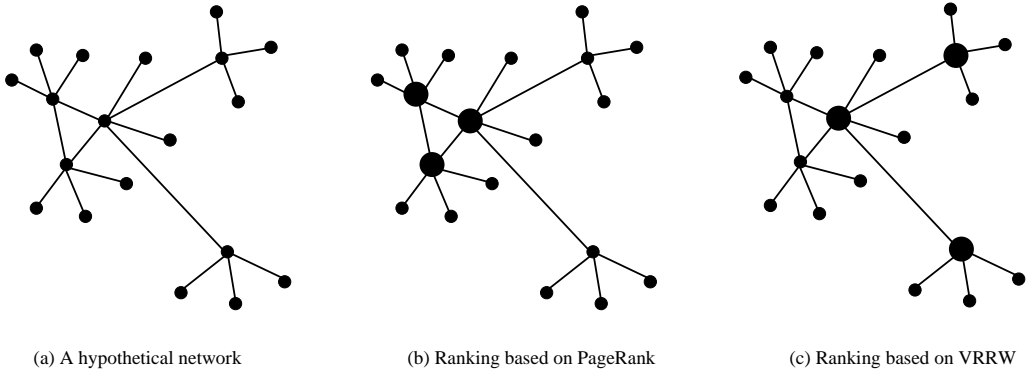}
 \caption{(a) A hypothetical network; (b) top three vertices found using PageRank are highlighted with large circles, and (c) top three nodes found using
VRRW are highlighted with large circles. This example is taken from  \cite{discern}.}\label{example}
\end{figure}

\subsubsection{PageRank-based Selection} This approach leverages the well-known PageRank algorithm to select the top $S$ solutions which are of high quality but which are as diversely separated in the network as possible. We adopt Vertex Reinforced Random Walk (VRRW), a time heterogeneous random walk process used by  \cite{Mei:2010}. Let us assume that we construct a network, where vertices are the base solutions and the weight of an edge connecting two vertices $i$ and $j$ indicates the pair-wise diversity ($1-\mathbb{Q}$) between corresponding base solutions. Unlike in traditional PageRank where the transition probabilities remain static throughout the iterations, in VRRW they change over time/iteration based on the following equation:
\begin{equation}
 p_T(i,j)=(1-\lambda)  \cdot  p^*(j)+\lambda  \cdot  \frac{p_0(i,j)  \cdot  p_T(j)}{D_T(i)}
\end{equation}
where
\begin{equation}
 D_T(i)=\sum_{k\in V}p_0(i,j)p_T(k)
\end{equation}
Here, $p_T(i,j)$ is the transition probability from vertex $i$ to vertex $j$ at time $T$, $p^*(j)$ is a distribution which represents the prior
preference of visiting vertex $j$, $p_0(i, j)$ is the ``organic'' transition probability prior to any reinforcement. $\lambda$ is set to $0.9$ as suggested by \cite{discern}. $p_T(k)$ denotes the probability that the walk is at vertex $k$ at time $T$: $p_T(k)=\sum_{(n,k)\in E} p_T(n,k)p_{T-1}(n)$. We set  $p^*(j)=Quality(C_j,\Gamma)$ and $p_0(i,j)$ as follows:
\begin{equation}
 p_0(i,j)=\left\{\begin{array}{cll}
	\alpha \cdot \frac{(1-\mathbb{Q}(\mathbb{C}_i,\mathbb{C}_j))}{Weighted\_deg(\mathbb{C}_i)} & if\ i\neq j \\
        1-\alpha, & otherwise\\
	\end{array}\right.
\end{equation}
where $Weighted\_deg(\mathbb{C}_i)$ is the weighted degree of vertex $i$, representing community structure $\mathbb{C}_i$. A schematic diagram of the VRRW process compared to PageRank is presented in Figure \ref{example}\footnote{Readers are encouraged to read the details in \cite{Mei:2010}.}. In this strategy, vertices are ranked based on VRRW score at the stationary state. Then we collect top $S$ high ranked vertices, which in turn produces $S$ high quality base communities which are diversely separated in the network.

\subsection{Experimental Results}
\begin{figure}[!t]
 \centering
 \includegraphics[width=\columnwidth]{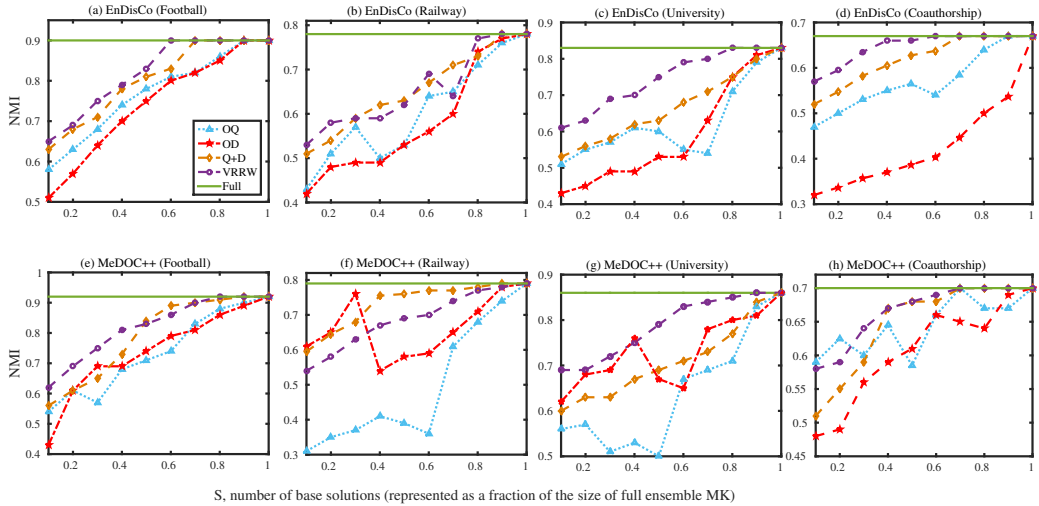}
 \caption{Comparison of the performance of different selection strategies: OQ: only quality, OD: only diversity, Q+D: combining quality and diversity with $\alpha=0.5$, VRRW: vertex reinforced random walk, with the variation of the size of the base solutions for disjoint community detection. We compare the performance of these strategies with the full ensemble (Full) represented by the solid green line. }\label{disjoint_strategy}
\end{figure}

\begin{figure}[!t]
 \centering
 \includegraphics[width=\columnwidth]{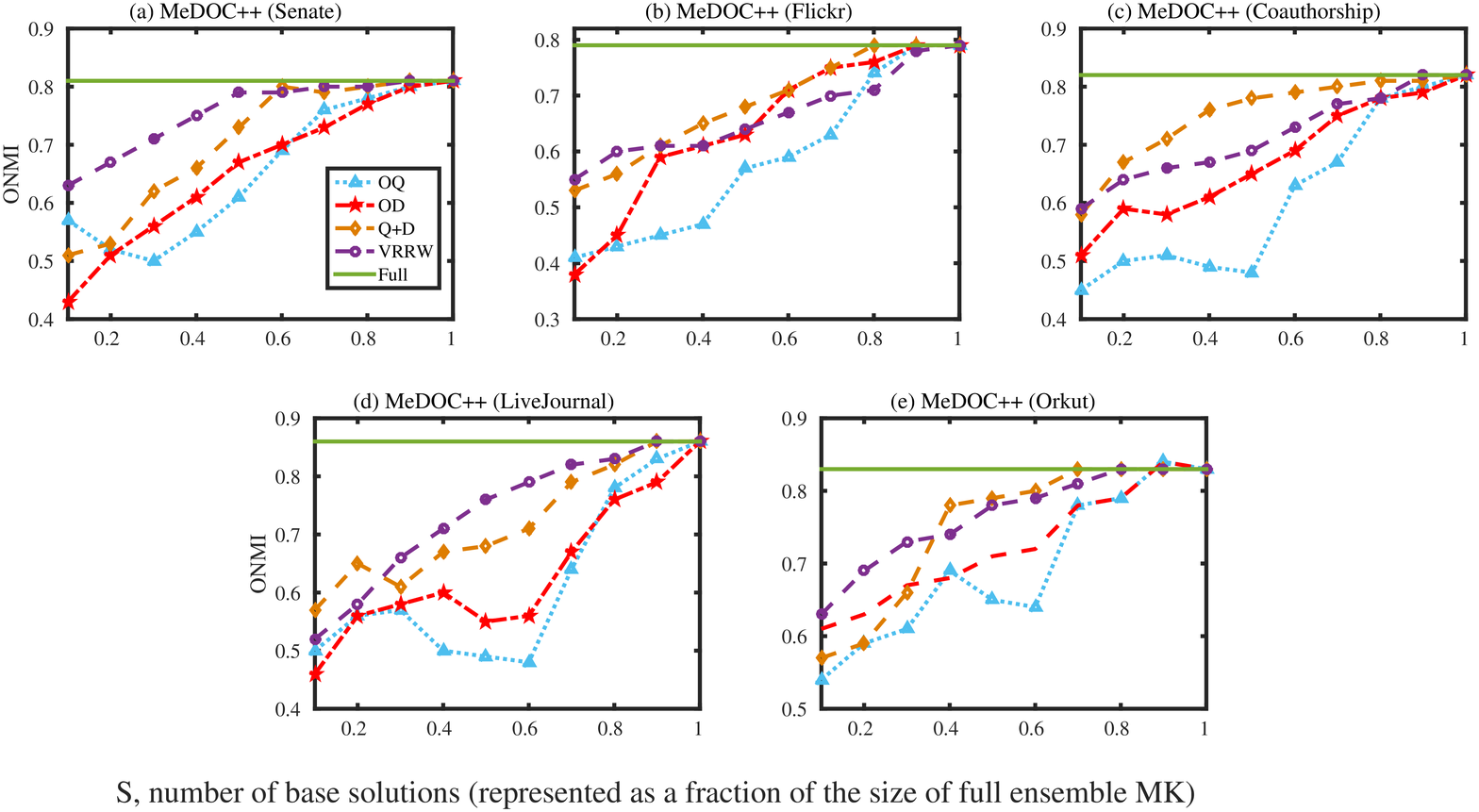}
 \caption{Comparison of the performance of different selection strategies: OQ: only quality, OD: only diversity, Q+D: combining quality and diversity with $\alpha=0.5$, VRRW: vertex reinforced random walk, with the variation of the size of the base solutions for overlapping community detection. We compare the performance of these strategies with the full ensemble (Full) represented by the solid green line. }\label{overlapping_strategy}
\end{figure}

To evaluate the ensemble selection strategies, we apply each strategy on all the datasets separately for disjoint and overlapping community detections. The results are reported by averaging the values after ten independent runs. In Figures \ref{disjoint_strategy} and \ref{overlapping_strategy} we plot the performance of different selection strategies as a function of ensemble size $S$ (where $S$ is represented as a fraction of the full ensemble set). We also show the results after considering all the base solutions in the ensemble set using the horizontal green line.

It is evident from both these results that the full ensemble is always best. However, we might achieve the same performance by selecting a subset of the base solutions. Both ``combining quality and diversity'' and ``VRRW'' strategies seem to be more consistent and achieve promising performance toward
our goal, that is to select smaller and better performing ensembles. Among them, the VRRW-based strategy tends to achieve the maximum accuracy with just 60-80\% size of the full ensemble. For the large networks such as Coauthorship network in Figure \ref{disjoint_strategy}(d), the separation of the performance is much prominent for \ENDISCO~and VRRW-based strategy reaches the maximum accuracy with 40\% of the full ensemble. The reason might be that 
this strategy explicitly seeks to remove redundant solutions and retains quality solutions at the same time.\\

\begin{figure}[!t]
 \centering
 \includegraphics[width=1.05\columnwidth]{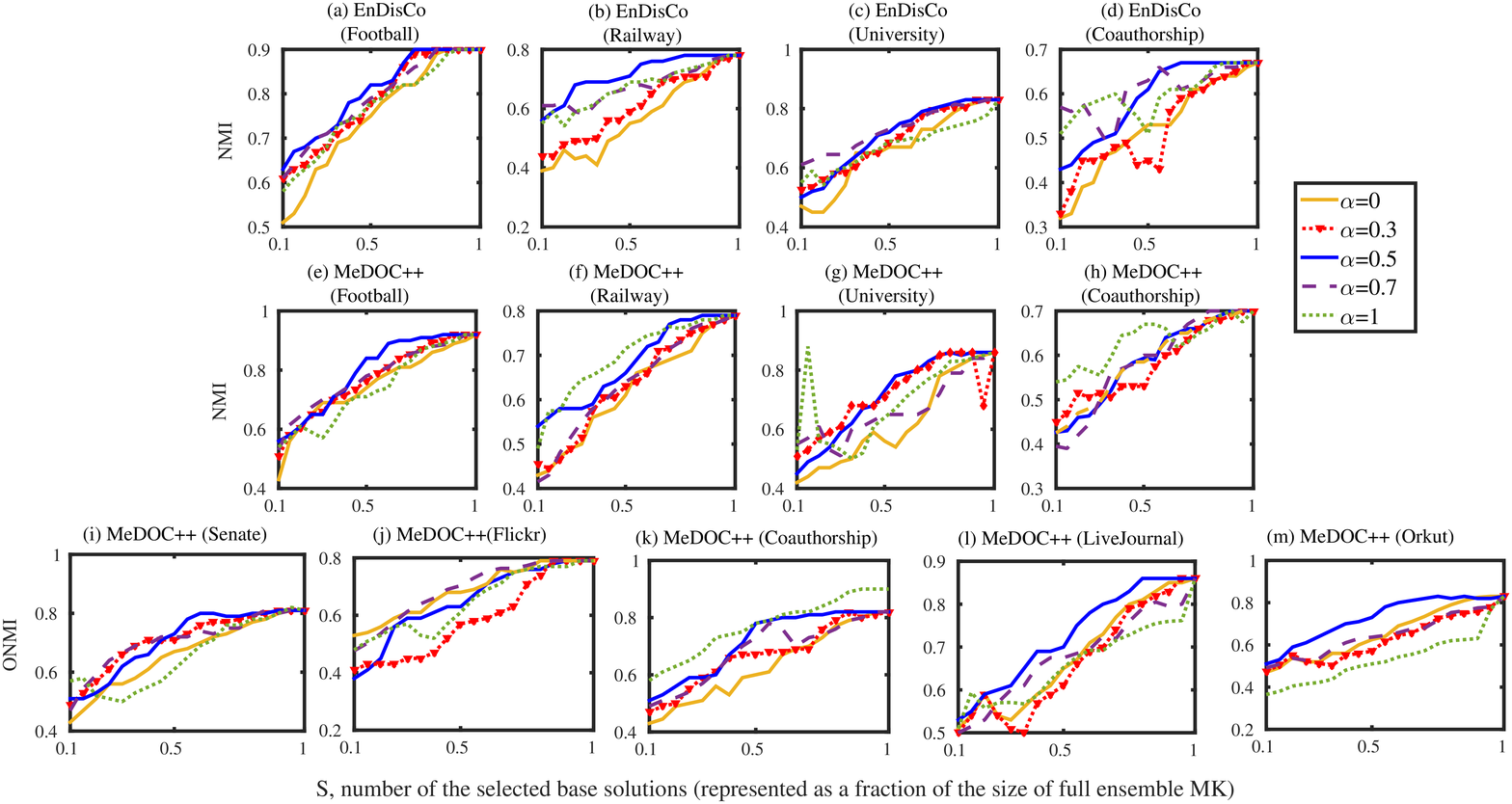}
 \caption{Sensitivity of the ``Combining quality and diversity'' strategy by varying the parameter $\alpha$ (0, 0.3, 0.5, 0.7, 1) for different sizes of the selected based solutions ($S$). $S$ is represented as a fraction of the size of full ensemble ($MK$), i.e., the size of the base solutions when all base solutions are taken. (a)-(h) Results of \ENDISCO~and \MEDOC~for disjoint community detection. (i)-(m) Results of \MEDOC~for overlapping community detection.}\label{alpha}
\end{figure}

\noindent\textbf{Sensitivity of $\alpha$ for ``combining quality and diversity'' strategy:} So far we conducted all the experiments with $\alpha=0.5$ for ``combining quality and diversity'' strategy. Here we examine how this strategy gets affected by varying the value of $\alpha$. We  experiment with a variety of $\alpha$ values including $0,0.1\cdots,0.9,1$ and compare the results with ``only quality'' ($\alpha=1$) and ``only diversity'' ($\alpha=0$). The smaller values of $\alpha$ takes ``diversity'' into account; whereas larger values prefer ``quality''. In Figure \ref{alpha}, we present the results of \ENDISCO~ and \MEDOC~for both disjoint and overlapping community detection by varying the size of the selected base solutions and five different values of $\alpha$ ($0, 0.3, 0.5, 0.7, 1$). Note that each accuracy value shown here is the average of ten runs. In general, we observe that assigning very high or very low values to $\alpha$ might be beneficial for some cases, but in general the result is more stable and robust for $\alpha=0.5$. We also observe that the accuracy never decreases with the increase of $S$ and gets saturated quickly for $\alpha=0.5$.

\section{Other Implications of MeDOC$++$}\label{sec:implication}
In this section, we present two other useful capabilities of \MEDOC. We show that \MEDOC~can be used to explore the core-periphery structure of communities in a network and to detect stable communities from dynamic networks. 

\subsection{Exploring Community-centric Core-periphery Structure}
The association values of individual vertices in a community obtained from \MEDOC~provide additional information about the membership strength in that community. This phenomenon might be related to the core-periphery organization \cite{PhysRevE.91.032803,ShaiCarmi07032007} of the induced subgraph composed of the vertices in a community. We hypothesize that the more the association of a vertex in a community, the more the vertex is likely to be a core part of the community. To verify this, we first explore the core-periphery structure (also known as k-core decomposition \cite{ShaiCarmi07032007}) of each community in a network.\\

\noindent{\bf Core-periphery organization:} For each community detected by \MEDOC, we start by creating the induced subgraph composed of vertices in that community. We then recursively remove vertices which have degree one until no such degree-one vertices remain in the network. The removed vertices form the 1-shell of the network ($k^s$
-shell index $k^s=1$). Similarly, we obtain $2$-shell by recursively removing degree-two vertices from the resultant network obtained from the last stage. We keep on increasing the value of $k$ until all the vertices are assigned to at least one shell. Then the $k^s$-core is formed by taking the union of all the shells with index greater than or equal to $k^s$.

\begin{figure}[!t]
 \centering
 \includegraphics[width=\columnwidth]{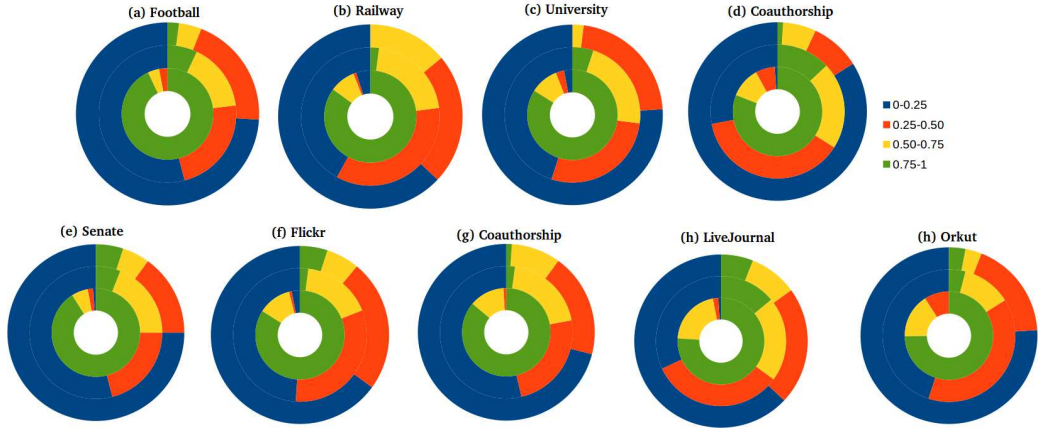}
 \caption{Core-periphery organization of the disjoint (top panel) and overlapping (bottom panel) communities detected by \MEDOC~ and its relation with the association values. The colors represent different ranges of association values and the area covered by each colored region in each $k^s$-shell denotes the proportion of vertices with the corresponding association values occupied in that shell. The innermost shell is the core region and the outermost shell is the periphery region. For better visualization, we divide the total number of shells identified from the communities in each network into three broad shells and the range of association values into four buckets; thus the core-periphery structure in each network has three concentric layers and each layer is divided into four association ranges. }\label{disjoint_core}
\end{figure}

The idea is to show how the association value of a vertex to a community is related to its community-centric shell-index\footnote{Note that the average core-periphery structure of communities in a network is different from the core-periphery structure of a network. Here we are interested in the former case.}. For each network, we first detect the community using \MEDOC~and measure the association values of vertices. Then, for each detected community, we extract the core-periphery structure.   In Figure \ref{disjoint_core}, we present the correlation between association values of vertices with their positions in the core-periphery structure. For better visualization, we divided the total number of shells into three broad shells, and the range of association value into four buckets. We note that the inner core is highly dominated by the vertices with association value ranging $0.75-1$, whereas the outermost layer is dominated by vertices having association value $0-0.25$. Further, we observe that as we move from core to periphery, vertices with low association score start dominating. This result indicates that the association value derived from \MEDOC~has high correlation with the core-periphery origination of the community structure.

\subsection{Discovering Stable Communities in Dynamic Networks}

Most real-world networks are dynamic -- either vertices or edges get added to the network or get deleted over time. As a result, the community structure of the network might change over time. \cite{Xiao-qiang:1977} argued that despite the change in community structure, there are some stable communities that persist.
Such stable communities may be important for many reasons. 
For instance, in an election campaign, vertices in stable communities might be individuals who are strong supporters of a particular candidate. In human interaction networks, the stable communities might be the groups of  individuals who are close friends or family members. For instance, there are a number of efforts
\cite{eagle2007isn}
in which researchers have provided subjects with cell phones (or cell phone apps), and a temporal human interaction network is built by identifying mobile phones that are in close proximity with one another during a given time window. In such cases,
communities that are stable between 8am and 6pm might represent people who work together, while communities that are stable during the 8pm to 6am time frame might represent families.
We hypothesize that the association values of individual vertices obtained via the \MEDOC~algorithm can discover stable communities in dynamic networks. In particular, vertices with an association value tending to $1$ might form stable communities in a network.

To this end, we collect two dynamic networks with unknown ground-truth community structure\footnote{\url{http://www.sociopatterns.org/datasets/}} \cite{tanmoy:jcs}: (i) {\em Hypertext 2009 contact network}: this is a dynamic network of face-to-face proximity over about 2.5 days of 110 attendees who attended ACM Hypertext 2009 conference, (ii) {\em School contact network}: this dataset contains the temporal network of contacts between 298 students over 4 days in a high school in Marseilles, France. We divide each network into five subnetworks based on the equal division of the temporal dimension in such a way that number of vertices is same in each subnetwork. However the edge connectivity might vary across five time windows. 

In each time window $t_i$, we run \MEDOC~ on the corresponding network. We allow \MEDOC~to detect communities with the default parameter setting. Once we obtain the association matrix $\mathbb{A}$ (in Step \ref{algo2:asso} of Algorithm \ref{meclud}), we considered {\em only} those vertices whose association score is exactly $1$ and group them accordingly (resulting disjoint stable community structure).  We believe that these vertices form the stable communities. We repeat the same procedure for each temporal network separately and detect the stable community structure. Then we measure the similarity (based on NMI and ARI) between stable community structures obtained from the temporal networks in two consecutive time windows ($t_i$ and $t_{i+1}$) , where $i$ ranges from $1$ to $4$) of each dynamic network. In Table \ref{tem}, we observe that the temporal similarity between two consecutive stable communities is quite high,  and it remains almost same over time. This indicates that the stable community structure of a network do not vary much over the successive time periods, and \MEDOC, quite efficiently, detects such stable communities from dynamic networks.

\begin{table}
 \centering
 \caption{Similarity of the stable communities of temporal networks in two consecutive time stamps obtained from \MEDOC~for two dynamic networks (Hypertext and school contact networ).}\label{tem}
 \begin{tabular}{|c|c|c|c|c|c|}
 \hline
  Network & Measure & $t_1 - t_2$ & $t_2 - t_3$ & $t_3 - t_4$ & $t_4 - t_5$\\\hline
  \multirow{2}{*}{Hypertext} & NMI & 0.79 & 0.76 & 0.80 & 0.82 \\\cline{2-6}
                             & ARI & 0.81 & 0.78 & 0.79 & 0.83\\\hline
   \multirow{2}{*}{School} & NMI & 0.76 & 0.79 & 0.78 & 0.80\\\cline{2-6}
                           & ARI & 0.75 & 0.76 & 0.82 & 0.83\\\hline  
 \end{tabular}
\end{table}

\section{Handling Degeneracy of Solutions}\label{sec:degeneracy}
Most community finding algorithms suffer from the well-known ``degeneracy of solutions'' problem \cite{good2010performance} which states that these algorithms typically produce exponentially many solutions with (nearly-)similar optimal value of the objective function (such as modularity); however the solutions may be structurally distinct from each other. 

\if{0}
Figure \ref{example} shows a small example of how \INF~ produces many outputs for different vertex orderings of Football network. 

\begin{figure}[!h]
 \centering
 \scalebox{0.2}{
 \includegraphics{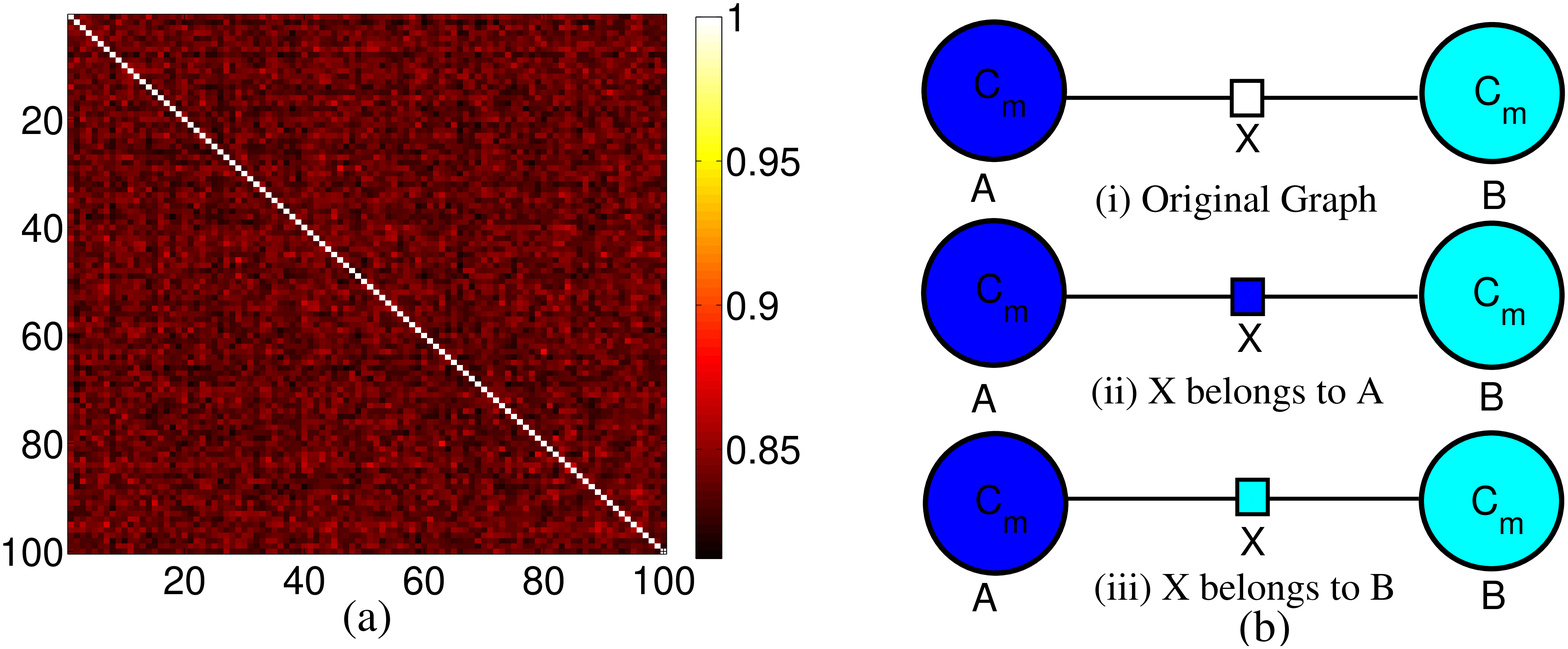}}
 \caption{(a) Similarity between pair-wise community structures (based on NMI \cite{danon2005ccs}) after running Infomap algorithm on $100$ different vertex orderings of Football network. (b) A schematic network consisting of two cliques $A$ and $B$ of size $m$ (representing two communities) connected by a bridging vertex $X$. Assigning $X$ to either $A$ or $B$ yields the same value of the optimization metrics (such as modularity, conductance).}\label{example}
\end{figure}
\fi

\begin{figure}[!t]
\centering
 \scalebox{0.28}{
 \centering
  \includegraphics{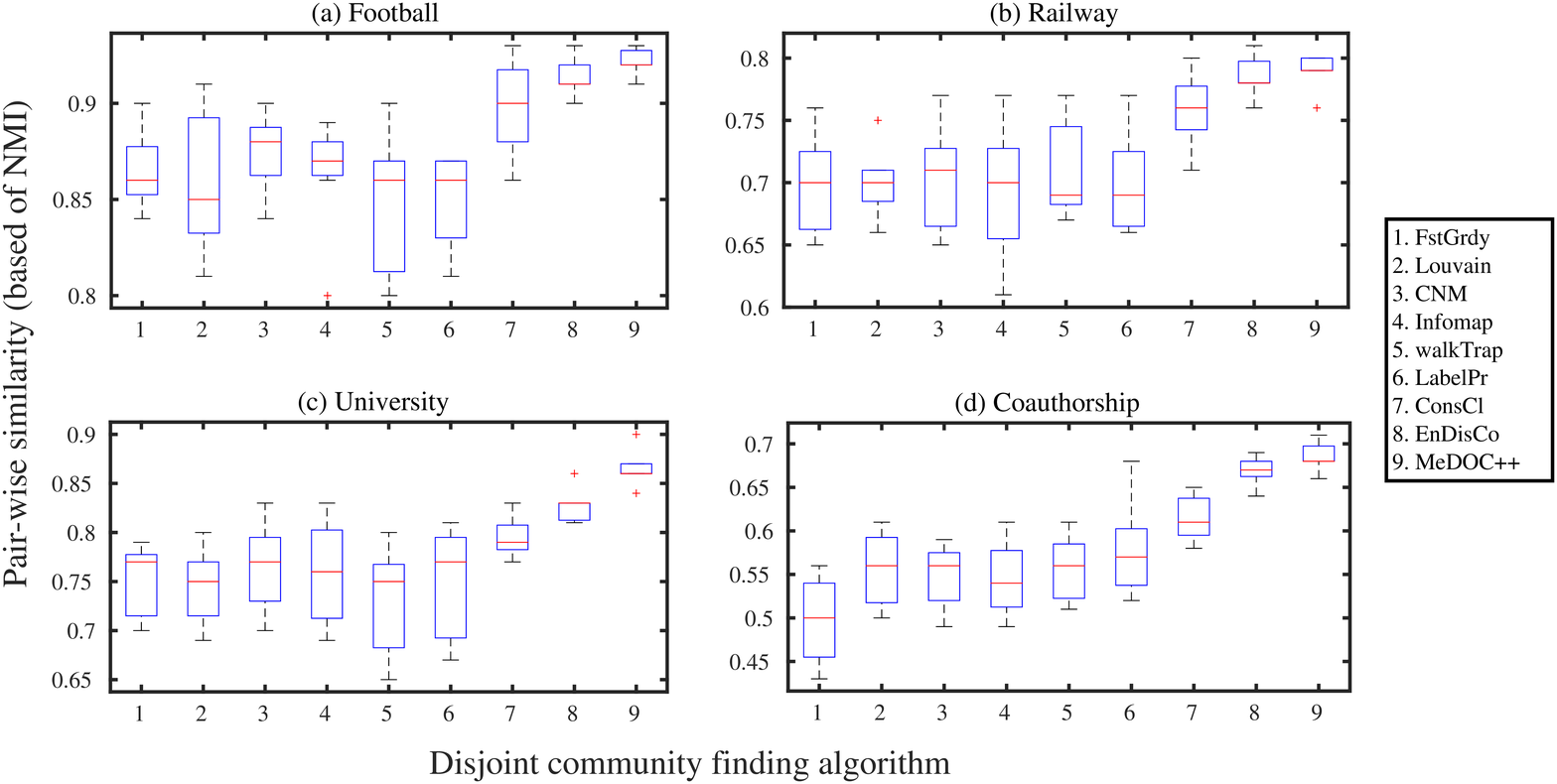}
  }
  \caption{Box plots indicating the variation of the solutions obtained from the disjoint community finding algorithms on real-world networks.  }\label{boxplotdisjoint}
\end{figure}

\begin{figure}[!t]
\centering
 \scalebox{0.28}{
 \centering
  \includegraphics{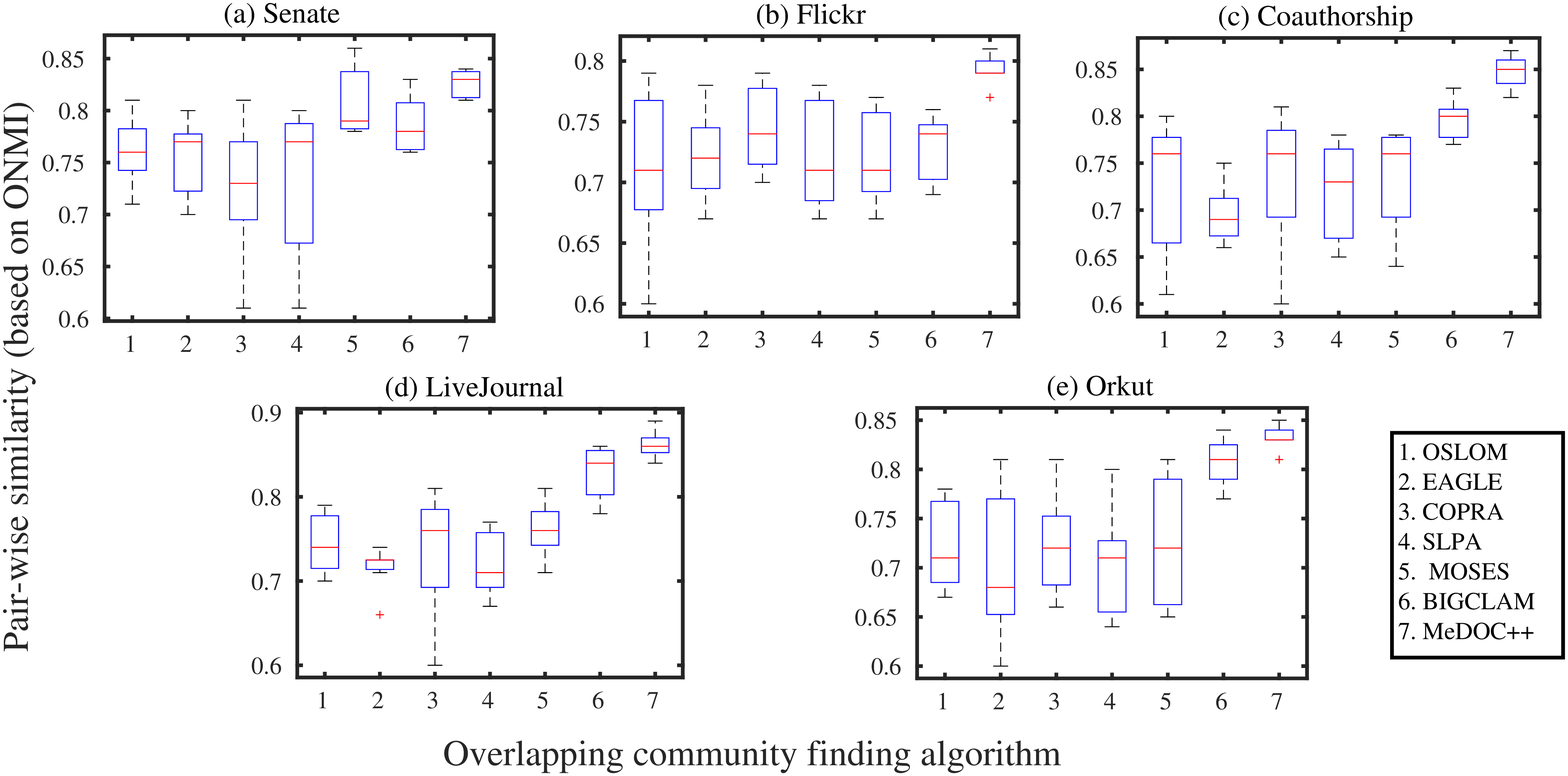}
  }
  \caption{Box plots indicating the variation of the solutions obtained from the overlapping community finding algorithms on real-world networks.  }\label{boxplotoverlapping}
\end{figure}

We hypothesize that since ensemble algorithms combine multiple views of the underlying community structure of a network, they might suffer less from the problem of degeneracy of solutions. We test this by considering the real-world networks, and running each competing algorithms on $100$ different vertex orderings of each network. We then measure the pair-wise similarity of the solutions obtained from each algorithm. The box plots in Figures \ref{boxplotdisjoint} and \ref{boxplotoverlapping} show the variation of the solutions for all the competing algorithms on both disjoint and overlapping community detections respectively. We observe that the median similarity is high with \ENDISCO~and \MEDOC~ and the variation is significantly less. In fact, all the ensemble algorithms, i.e., \ENDISCO, \MEDOC~and \CC~show low variance. These results suggest that ensemble based algorithms always provide more robust results than standalone algorithms and alleviate the problem of degeneracy of solutions.

\section{Runtime Analysis}\label{sec:speed}
\begin{figure}[!t]
\centering
 \scalebox{0.25}{
  \includegraphics{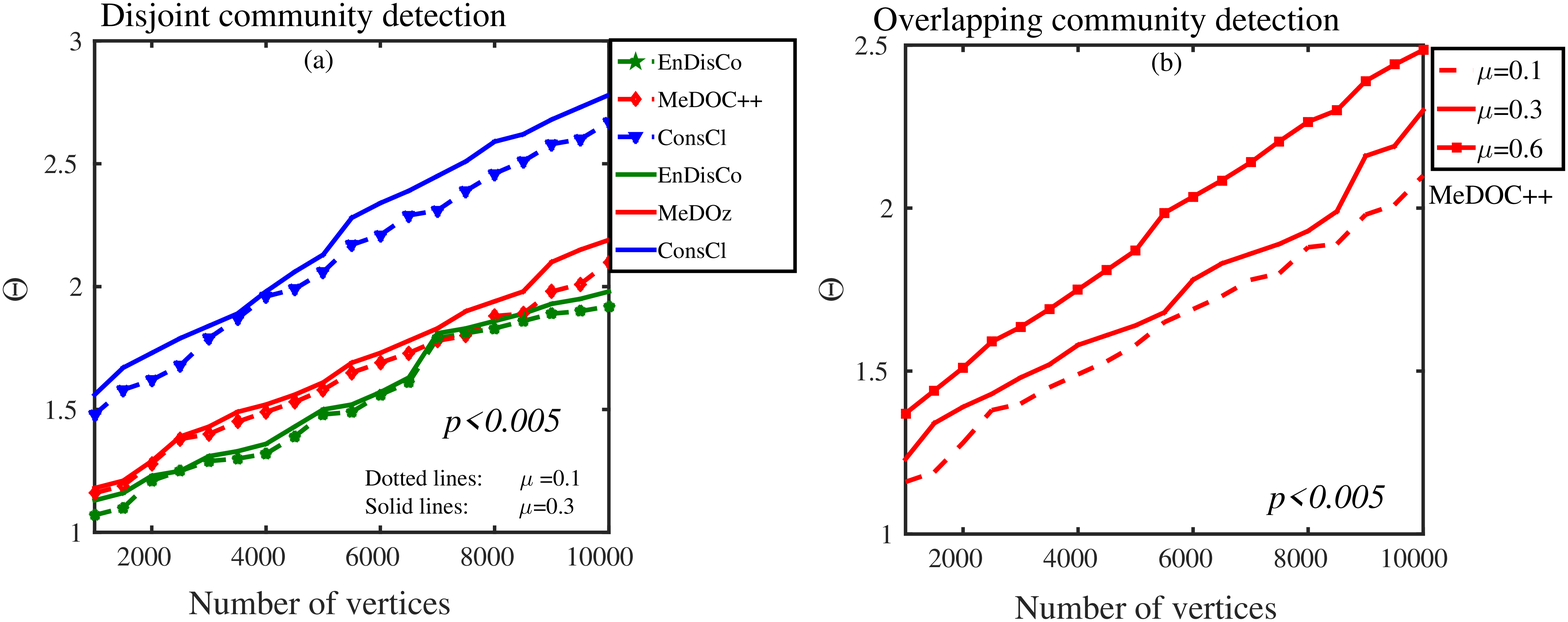}}
  \caption{The value of $\Theta$ w.r.t. the increase of vertices in LFR networks. \ENDISCO~and \MEDOC~ are compared with \CC. The results are statistically significant (since there are multiple curves, we report the range of $p$-values).}\label{runtime}
   
\end{figure}

Since ensemble approaches require the running all base algorithms (which may be parallelized), one cannot expect ensemble methods to be faster than standalone approaches. However, our proposed ensemble frameworks are much faster than existing ensemble approaches such as consensus clustering. To show this, for each ensemble algorithm, we report $\Theta$,
the ratio between the runtime of each ensemble approach and the sum of runtimes of all base algorithms, with increasing number of vertices in LFR. We vary the number of edges of LFR by changing $\mu$ from $0.1$ to $0.3$. Figure \ref{runtime} shows that  our algorithms are much faster than consensus clustering. We further report the results of \MEDOC~for overlapping community detection which is almost same as that of disjoint case since it does not require additional steps apart from computing the threshold.

\section{Conclusion}
In this paper, we proposed two general frameworks for ensemble community detection. \ENDISCO~identifies disjoint community structures, while \MEDOC~detects both disjoint, overlapping and fuzzy community structures. 
\MEDOC\ is the first ever ensemble overlapping (and fuzzy) community detection algorithms in the literature.

We tested our algorithms on both synthetic data using the LFR benchmark as well as with several real-world datasets that have an associated ground-truth community structure. We showed that both \ENDISCO~and \MEDOC~are more accurate than existing standalone community detection algorithms (though of course, \ENDISCO~and \MEDOC~leverage them by using the known community detection algorithms in the ensembles).
We further showed that for disjoint community detection problems, \ENDISCO~and \MEDOC~both beat the best performing existing disjoint ensemble method called consensus clustering \cite{lanc12consensus} -- both in terms of accuracy  and run-time. To our knowledge, \MEDOC~is the first ensemble algorithm for overlapping and fuzzy community detection that we have seen in the literature. 

We further presented extensive analysis of how to select a subset of solutions to obtain near-optimal accuracy. The association values of vertices obtained from \MEDOC~help us exploring the core-periphery structure of the communities in a network and detecting the stable communities in dynamic networks. We showed that ensemble approaches  generally reduce the effect of degeneracy of solution significantly.

\if{0}
In this paper, we proposed two frameworks to aggregate multiple community structures obtained by running different disjoint CD algorithms. Both the frameworks turned out to be superior than past non-ensemble based approaches as well as a recently proposed ensemble approach. We showed how one can leverage disjoint community information  to discover the overlap in the community structure. We presented the suitable functions needed in the process of ensemble and showed an automated way of selecting the threshold. In that sense, our algorithms do not require manual intervention for parameter tunning. 
As it turned out, the choice of base algorithm has no major impact on the clustering quality, but it does in selecting an algorithm for re-clustering.

 In future, we would like to develop theoretical explanation to justify the superiority of ensemble approaches compared to the discrete models.
In traditional ensemble-based classification problem, it has already been shown that an aggregation of several weak base algorithms performs well even though each weak algorithm is just merely better than the random guessing,  i.e., the probability of correct classification is $\frac{1}{|C|} + \epsilon$, where $\epsilon$ is a small non-zero  value and $|C|$ is the number possible classes   \cite{Freund:1995:BWL:220262.220446,Schapire:1990:SWL:83637.83645}. In case of community detection, one can think of weak base algorithms that find the correct community for a vertex in the probability of $\frac{1}{|C|} + \epsilon$ (boosting condition).  After a series of boosting procedures, e.g., boosting by a majority voting, the small $\epsilon$ value would help the probability that the community structure is correctly identified converging to a much larger value than $\frac{1}{|C|} + \epsilon$.  Of course, there exists a certain probability that the majority vote may not work correctly. If the boosting condition is guaranteed for every vertex and the number of weak algorithms is enough, taking a majority vote may produce stable results. In many practical cases, however, we found that the community detection for a subset of vertices is significantly inaccurate by a majority of weak algorithms, and thus the boosting condition might not be guaranteed for all vertices. We leave this line of research as future agenda.
\fi

\section*{References}
\bibliographystyle{model1-num-names}
\bibliography{ref,ref1,sigproc}







\end{document}